\documentclass[twocolumn]{aastex631}

\newcommand{\HI}{H\,\textsc{i} }
\newcommand{\kms}{km.s^{-1}}

\usepackage{amsmath}
\usepackage{cases}
\usepackage{multirow}
\usepackage{siunitx}
\usepackage[version=4]{mhchem}
\usepackage{hyperref}

\shorttitle{Structure decomposition algorithm}
\shortauthors{Liu and Du}
\graphicspath{{./}{figures/}}

\begin{document}

\title{Atomic and molecular gas in the Milky Way. I. Structure decomposition}

\correspondingauthor{Fujun Du}
\email{fjdu@pmo.ac.cn}

\author[0000-0002-0409-7466]{Xin Liu}
\affiliation{Purple Mountain Observatory, Chinese Academy of Sciences, Nanjing 210023, China}
\affiliation{School of Astronomy and Space Science, University of Science and Technology of China, Hefei 230026, China}

\author[0000-0002-7489-0179]{Fujun Du}
\affiliation{Purple Mountain Observatory, Chinese Academy of Sciences, Nanjing 210023, China}
\affiliation{School of Astronomy and Space Science, University of Science and Technology of China, Hefei 230026, China}



\begin{abstract}
We present GDCluster, a fully automated algorithm for decomposing spectral-line datacube of interstellar gas into coherent structures.
Assuming a multi-Gaussian nature of observed spectra, GDCluster employs and augments the derivative spectroscopy technique for precise parameter estimation, incorporates spatial-continuity constraints during spectral fitting, and extends these constraints to spatial clustering.
This approach effectively resolves velocity blending structures in PPV space—particularly critical for ubiquitous \HI spectra where emissions from multiple phases are severely blended.
Applied to the all-sky HI4PI data, a \SI{10}{\degree}$\times$\SI{10}{\degree} CRAFTS survey region, and a \SI{45}{\degree}$\times$\SI{10}{\degree} MWISP survey region, GDCluster extracts \num[group-separator={,}]{45299}, \num[group-separator={,}]{2247}, and \num[group-separator={,}]{47119} structures in \HI and $\ce{CO}$ (1--0), respectively.
Comparative analyses demonstrate GDCluster's superiority over DBSCAN in separating overlapping spectra with complex velocity components.
\end{abstract}



\section{Introduction}
Molecular clouds (MCs) are the birthplaces of stars. Understanding their formation is a crucial first step toward studying the life cycle of matter in galaxies. In the local Universe, molecular hydrogen (H$_2$) primarily forms from atomic hydrogen (H\,\textsc{i}) on dust grain surfaces. Consequently, the kinematic and morphological relationship between atomic and molecular gas serves as a key indicator of MC formation.

The \HI gas in the Milky Way can be divided into three distinct phases, with temperatures ranging from a few tens to thousands of Kelvin \citep{2003ApJ...587..278W}. While the low-density warm neutral medium (WNM) is associated with large-scale structures \citep{2006Sci...312.1773L, 2008A&A...487..951K}, only the cold neutral medium (CNM) provides sufficient ultraviolet shielding and low temperatures necessary for MC formation. Due to its high optical depth and relatively low temperature, CNM structures are more readily observed through absorption \citep{2003ApJ...586.1067H}, enabling detailed morphological comparisons with other tracers. Numerous studies have demonstrated that filamentary CNM structures are aligned with magnetic fields \citep{2014ApJ...789...82C, 2016ApJ...821..117K}, dust \citep{2019ApJ...874..171C}, and MCs \citep{1978ApJ...224..132B, 2020A&A...634A.139W, 2023ApJ...948L..17L, 2024ApJ...973L..27S}, whereas the structural characteristics of the WNM remain comparatively poorly understood. As one of the two thermally stable \HI phases alongside the CNM, the smoothly distributed WNM constitutes approximately 52\% of the \HI mass \citep{2018ApJS..238...14M} and shows little correlation with CNM structures \citep{2018A&A...619A..58K}. Although high-sensitivity observations against bright background sources can resolve CNM structures via absorption features, they currently suffer from limited spatial coverage \citep{2023ARA&A..61...19M}, which confines the scope of detectable CNM structures. In contrast, more general \HI 21 cm surveys offer high angular resolution \citep{2018ApJS..234....2P} and full-sky coverage \citep{2009ApJS..181..398M, 2016A&A...585A..41W, 2016A&A...594A.116H}.

A natural approach to studying the CNM via \HI emission spectra is to decompose the spectra into distinct components \citep{1985ApJ...289..792K}. By assuming a Gaussian distribution for the random radial velocities of individual \HI components \citep{1955ApJ...121..569H}, each line of sight (LoS) can be modeled as a sum of Gaussian components, characterized by a central velocity ($v_{\mathrm{c}}$), full width at half maximum (FWHM), and peak brightness temperature ($T_{\mathrm{peak}}$). Although \HI emission profiles are often complex—affected by velocity crowding and optical depth effects \citep{1990ARA&A..28..215D}, leading to deviations from Gaussian shapes, our view is that this Gaussian-decomposition approach is currently the best practical approach for extracting useful physical parameters from \HI spectral cubes.

Under these considerations, \citet{2000A&A...364...83H} and \citet{2006BaltA..15..413H} adopted a minimalist approach, employing as few Gaussian components as necessary while incorporating spatial information from neighboring sky positions to decompose the Leiden/Dwingeloo Survey data. \citet{2015AJ....149..138L} applied machine learning techniques to smooth the \HI spectra and introduced derivative spectroscopy to provide improved initial guesses for subsequent fitting routines.
\citet{2019A&A...628A..78R} enhanced the performance of this method by incorporating spatial coherence into the fitting process.
\citet{2017ApJ...834...57M} used a hierarchical algorithm to cluster Gaussian components into molecular clouds and recovered 98\% of the $\ce{^{12}CO}$ emission within Galactic latitude $b=\pm\SI{5}{\degree}$.
\citet{2019A&A...626A.101M} introduced spatial continuity of Gaussian parameters as constraints, deriving initial guesses at varying resolutions. \citet{2019MNRAS.485.2457H} used a semi-automatic fitting algorithm to decompose spectra and developed a hierarchical clustering method to group components into coherent structures. \citet{2020ApJ...902..120P} further modified the hierarchical clustering algorithm proposed by \citet{2017ApJ...834...57M} to identify \HI clouds at high Galactic latitudes. \citet{2024AJ....167..220Z} combined the multi-Gaussian fitting method of \citet{2019A&A...628A..78R} with the clustering approach of \citet{2019MNRAS.485.2457H} to analyze $\ce{^{13}CO}$ data in the Cygnus region. \citet{2024RAA....24k5005F} introduced a novel clustering algorithm based on graph theory to group Gaussian components into structures.

In this work, we present a fully automated algorithm, Gaussian Decomposition and Coherent clustering (GDCluster), for decomposing spectral-line data into coherent structures. The algorithm is specifically designed for separating gas with highly complex velocity components. Starting from the multi-Gaussian model, we refine the derivative spectroscopy technique to generate accurate initial estimates of Gaussian parameters.  Spatial continuity is leveraged by incorporating reliable initial guesses from adjacent LoSs, to improve the quality of the initial values.
To the best of our knowledge, our work is the first attempt to decompose the all-sky \HI emission into distinct \emph{structures}.  As mentioned before, only CNM structures exhibit morphological correspondence with molecular gas, yet they are often obscured by the WNM.  Structure decomposition is thus a critical prerequisite for investigating the H$\,\textsc{i}$-to-H$_2$ transition. Separating different phases of \HI enables detailed studies of phase transitions and interactions among the CNM, WNM, and the unstable neutral medium (UNM) \citep{2001ApJ...551L.105H}.
A single clustering criterion, namely velocity continuity, is employed in the clustering process, which ensures that the algorithm is readily interpretable. 
This simplification ensures clarity in the clustering process.
Although developed primarily with \HI data in mind, the GDCluster algorithm is also applicable to other tracers whose spectra can be approximated as a superposition of multiple Gaussians, such as $\ce{CO}$. 
Since $\ce{CO}$ structures can serve as indicators of foreground CNM structures \citep{2003ApJ...585..823L}, an algorithm applicable to both \HI and $\ce{CO}$ data is more reasonable and convenient for further analysis.

In Section~\ref{sec::method}, we describe the proposed decomposition algorithm in detail. 
Section~\ref{sec::application} presents its application to survey data, including a comparative analysis with other alternative methods.
Section~\ref{sec::discussion} addresses the limitations, efficiency, and hyper-parameter dependence of GDCluster, as well as the definition of cloud structures, followed by a summary in Section~\ref{sec::summary}.

\section{Methodology}
\label{sec::method}
\begin{figure*}[!ht]
    \centering
	\includegraphics[width=1.\textwidth]{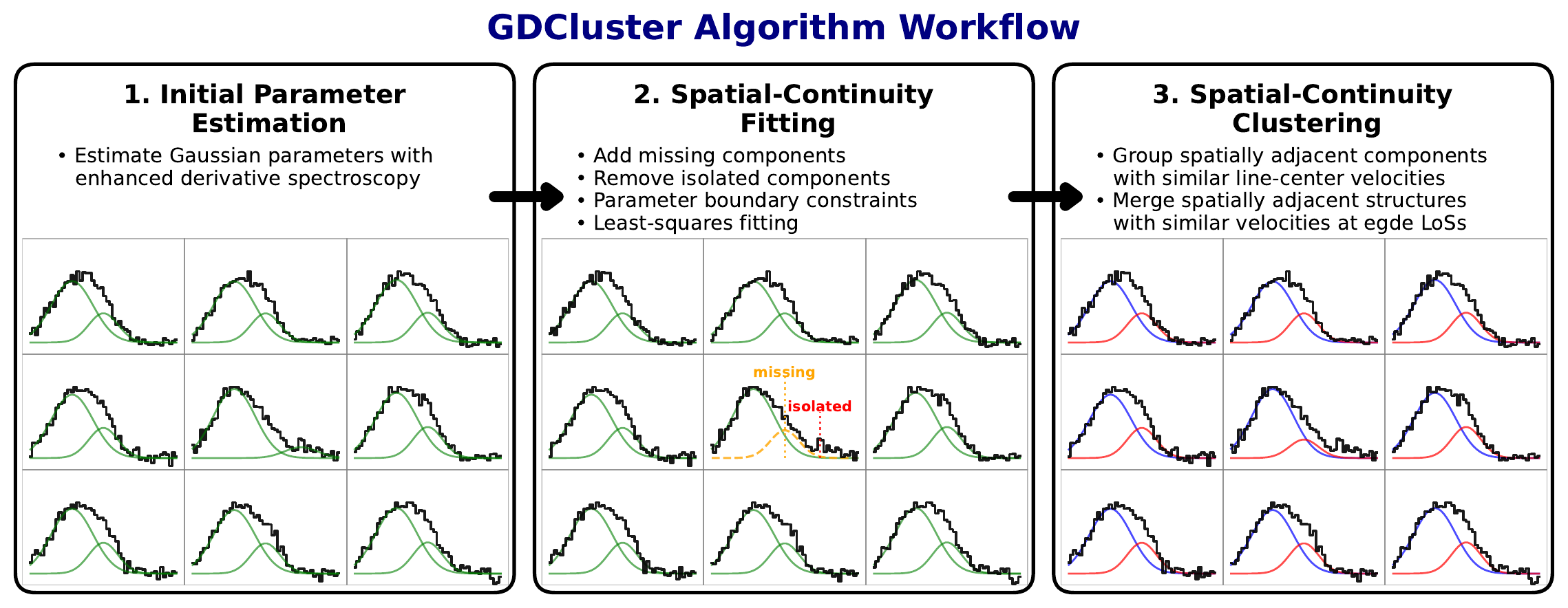}
	\caption{Workflow of the GDCluster algorithm. Black lines in each panel represent mock spectra from nine neighboring LoSs. Green lines in the first panel show the initially estimated Gaussian parameters. The central spectrum contains an isolated component (far right) that appears only in this spectrum, and lacks a component (center) that is present in all neighboring spectra. In the second panel, the isolated component is removed and the missing component is added, illustrated by the orange and red dashed lines, respectively. In the third panel, Gaussian components belonging to the same cluster are shown in the same color across different LoSs.}
	\label{fig::workflow}
\end{figure*}

The proposed algorithm comprises three main parts: initial parameter estimation using derivative spectroscopy, multi-Gaussian spectral fitting with spatial-continuity constraints, and clustering with the same constraints.
A rigorous mathematical analysis of the multi-Gaussian model’s derivatives yields credible initial parameter estimates.
These initial values are used to calculate the similarities between spatially neighboring Gaussian components, and each component is assigned a set of neighboring counterparts based on the spatial-continuity of the gas structure.
The multi-Gaussian parameters of each spectrum—initially derived in isolation—are then refined by leveraging spatial continuity: components detected in a given LoS but not in its adjacent spectra are discarded, while components present in neighboring LoSs but absent at the central LoS are added to the central spectrum. The detailed criterion is discussed in Section \ref{sec::fitting}.
Furthermore, initial values from neighboring positions act as constraints, forming parameter boundaries for the subsequent least-squares fitting. Finally, the fitted components are clustered into individual structures using the same spatial-continuity constraints.
Figure \ref{fig::workflow} presents the workflow of the GDCluster algorithm, together with its application to mock spectra from nine adjacent LoSs.

\subsection{Estimation of initial parameters using derivative spectroscopy}
\label{sec::init_guess}

Fitting spectra using the least-squares requires an initial set of model parameters, and parameters sufficiently close to the optimal values are essential for rapid and accurate convergence.
The derivative spectroscopy technique is widely used to resolve blended spectral lines in fields related to spectrometry \citep{grum_derivative_1972, BOSCHOJEDA20131}.
It can be understood as a high-pass filter that effectively suppresses broad spectral features \citep{1968ApOpt...7...61S}, which means that the larger-width parts of a spectrum decrease more rapidly, and narrow, weak, and previously hidden peaks progressively emerge.

\subsubsection{Derivative spectroscopy in previous works: single-Gaussian approximation}
\label{sec::ds_in_previous_work}
Consider a Gaussian line profile characterized by a line-center $\mu$, a standard deviation line-width $\sigma$, and a peak intensity $A$. Its intensity as a function of the radial velocity $v$ can be expressed as
\begin{equation}
	G(v; A,\,\mu,\,\sigma)=Ae^{-(v-\mu)^{2}/2\sigma^{2}}.
	\label{equ::one_gau}
\end{equation}
Calculating the higher-order derivatives of Equation \ref{equ::one_gau}, we can notice that the derivative values at the line-center are equal to zero for odd-order derivatives:
\begin{equation}
		G^{(n)}(\mu) = 0,\,n = 1,\,3,\,5\,...,
		\label{equ::odd_deri_constraints}
\end{equation}
while reach local maxima or minima for even-order derivatives.
This property forms the basis for using zero-crossings in odd-order derivatives to locate line-centers.
It is important to note, however, that satellite patterns in higher-order ($>1$) derivatives can also generate zero-crossings away from the line-centers.
These features can be suppressed by incorporating information from even-order derivatives:
\begin{equation}
	\begin{aligned}
		G''(\mu) &< 0,\\
		G^{(4)}(\mu) &> 0. \label{equ::even_deri_constraints}
	\end{aligned}
\end{equation}
In practice, derivative calculations are approximated through numerical differentiation.

After estimating $\mu$ via higher-order derivatives, the other two parameters $A$ and $\sigma$ can be determined from a combination of the original profile (i.e., the zeroth-order derivative) and the second-order derivative:
\begin{equation}
	\begin{aligned}
		G(\mu) &= A,\\
		G''(\mu) &= -\frac{A}{\sigma^{2}}. \label{equ::one_gau_2nd_deri}
	\end{aligned}
\end{equation}
This estimation method have already adopted by \citet{2015AJ....149..138L}, but the situation becomes significantly more complex when multiple Gaussian components are considered.

\subsubsection{Refinement of derivative spectroscopy under the multi-Gaussian assumption}
A multi-Gaussian spectrum is given by
\begin{equation}
	I(v) = \sum_{i=1}^{M} G(v; A_{i},\,\mu_{i},\,\sigma_{i}), \label{equ::multi_gau}
\end{equation}
assuming there are $M$ Gaussian components at one LoS.
At the line-center of the $i$th Gaussian (derived using Equation \ref{equ::odd_deri_constraints}) we have
\begin{equation}
	\begin{aligned}
		G(\mu_i) &= A_i + \sum_{j\ne i}^{M}A_j e^{-\frac{(\mu_{i}-\mu_{j})^{2}}{2\sigma_j^{2}}},\\
		G''(\mu_i) &= -\frac{A_i}{\sigma_{i}^{2}}+\sum_{j\ne i}^{M}\frac{A_j}{\sigma_j^2}\left[\frac{(\mu_i-\mu_j)^{2}}{\sigma_j^2}-1\right]e^{-\frac{(\mu_i-\mu_j)^{2}}{2\sigma_{j}^{2}}}, \label{equ::multi_gau_2nd_deri}
	\end{aligned}
\end{equation}
where all remaining Gaussians contribute to the intensity at the $i$th line-center, which makes the estimation of \{$A_{i},\,\sigma_{i}$\} using Equation \ref{equ::one_gau_2nd_deri} inaccurate, especially when the Gaussians are too close relative to their line-widths.
To improve the estimation of $A_i$ and $\sigma_i$, we incorporate the contributions from the rest Gaussians as a correction term $C$ and calculate $A_i$ and $\sigma_i$ as
\begin{equation}
	\begin{aligned}
		A_i &= G(\mu_i) - \sum_{j\ne i}^{M}C^{0}(\mu_i; A_{j}, \mu_{j}, \sigma_{j}),\\
		\sigma_i &=\sqrt{-\frac{A_i}{G''(\mu_i)-\sum_{j\ne i}^{M}C^{2}(\mu_i; A_{j}, \mu_{j}, \sigma_{j})}}, \label{equ::corrected_estimation}
	\end{aligned}
\end{equation}
where $C^{0}(\mu_i; A_{j}, \mu_{j}, \sigma_{j})$ and $C^{2}(\mu_i; A_{j}, \mu_{j}, \sigma_{j})$ represent the contribution of $j$th Gaussian at $v=\mu_i$ in $0$th- and $2$nd-order derivatives, respectively. 
Note that $C$ depends on the parameters of all remaining Gaussians, which are not available initially. To address this issue, we calculate $C$ iteratively: at the beginning, the correction terms are ignored, reducing Equations \ref{equ::corrected_estimation} to Equations \ref{equ::one_gau_2nd_deri}. This simplification provides initial estimations of \{$A,\,\mu,\,\sigma$\}, from which $C$ is computed. The correct terms along with initial estimations are subsequently used to refine \{$A,\,\mu,\,\sigma$\}, and the process is repeated iteratively. In our experiments on simulated multi-Gaussian profiles, \{$A,\,\mu,\,\sigma$\} converge rapidly to their true values within a few iterations.

The multi-Gaussian nature also affects the zero-crossings in odd-order derivatives and, consequently, the estimation of $\mu$. However, correcting $\mu$ is significantly more complex than correcting the other two parameters, especially in the presence of noise. Therefore, we leave the estimation of $\mu$ unmodified in the multi-Gaussian case.

\subsubsection{Noise estimation and intensity cutoffs}

In practice, noise distorts the Gaussian shapes of original spectra and dominates the higher-order derivatives \citep{butler_analysis_1970}, making the estimation method proposed in the previous section inapplicable on real data. Specifically, noise can create false zero-crossings in odd-order derivatives. \citet{ohaver_signal--noise_1981} highlighted the importance of smoothing in higher-order derivatives and provided the relative signal-to-noise ratio (SNR) enhancement as a function of the smoothing window width.
In GDCluster, we translate the smoothing window width to the minimum typical line-width ($\sigma_{\text{min}}$) of the Gaussian profiles, applying Gaussian smoothing to preserve narrow peaks as much as possible.

Since noise affects the entire structure decomposition process, a reliable noise estimation method is essential. In large-area surveys, noise statistics vary from region to region due to environmental changes during the observation schedule. Using a single root-mean-square (rms) noise intensity for all spectra may introduce bias. Therefore, a pixel-by-pixel noise estimation method is necessary when processing survey data. Traditional approaches mask channels with apparent signals and calculate the noise rms ($\alpha$) from the remaining channels. However, these methods heavily depend on the selection of signal-free channels, which may not always be available—especially in the case of pervasive \HI emission. Here we present a fully automatic noise estimation method, which utilizes higher-order derivatives to detrend signals and calculate the rms of the original spectrum as:
\begin{equation}
	\alpha_{0} =\alpha_{n,\,\mathrm{f}}\cdot\left[\sum_{m=0}^{n}\left(\frac{n!}{(n-m)!m!}\right)^{2}\right]^{-\frac{1}{2}}
	\label{equ::noise_estimation}
\end{equation}
where $\alpha_{n,\,\mathrm{f}}$ is the standard deviation of $n$th-order derivative (approximated using the forward difference method) of the original spectrum and $n = \lfloor\log_{2}(N)\rfloor$ with $N$ the number of channels.
A detailed description of the proposed noise estimation method is shown in Appendix \ref{apx::noise_esti}. We highly recommend this method, as it is applicable not only to Gaussian profiles but for nearly all types of astrophysical spectra. Furthermore, this method is insensitive to the background continuum and the completeness of baseline removal.

After determining the noise rms for each spectrum, we can add noise cutoffs into Equation \ref{equ::even_deri_constraints} and the original spectrum to filter out zero-crossings caused by satellite patterns and noise:
\begin{equation}
	\begin{aligned}
		G(\mu) &> \text{SNR0}\cdot\alpha_0,\\
		G''(\mu) &< \text{SNR2}\cdot\alpha_2,\\
		G^{(4)}(\mu) &> \text{SNR4}\cdot\alpha_4,
		\label{equ::even_deri_constraints_noise}
	\end{aligned}
\end{equation}
where the SNR for the 0th- and, 2nd-, and 4th-order derivatives are set manually (typically $\mathrm{SNR}0=9$, $\mathrm{SNR}2=2$, and $\mathrm{SNR}4=1$, as seen in Table \ref{tab::hyperparameters}), $\alpha_0$ can be obtained using Equations \ref{equ::noise_estimation}.
In GDCluster, derivatives are approximated using the central difference method. Accordingly, by applying the rule of error propagation, $\alpha_2$ and $\alpha_4$ are obtained as:
\begin{equation}
    \begin{aligned}
    \alpha_2 &= \frac{\sqrt{6}}{4}\cdot\alpha_0,\\
    \alpha_4 &=\frac{\sqrt{70}}{16}\cdot\alpha_0.
    \end{aligned}
\end{equation}
These cutoffs, along with Gaussian smoothing, substantially suppress spurious peaks.
As higher-order derivatives are computed, progressively narrower Gaussian components emerge; however, this occurs at the expense of increasingly pronounced noise. To balance these two issues, zero-crossings in the 1st- and 3rd-order derivatives are utilized to determine the positions of line-centers. 

\subsubsection{Estimation error of the line-center introduced by noise}
Additionally, in the presence of noise $z$, line-centers may shift away from their true positions due to the noisy values used in their derivation (first-order estimation):
\begin{equation}
	\mu = v_{\text{L}} + (v_{\text{R}}-v_{\text{L}}) \cdot \frac{f(v_{\text{L}})}{f(v_{\text{L}})-f(v_{\text{R}})},
\end{equation}
where \( f(v) \) is the odd-order derivatives of the noisy spectrum, \( v_{\text{L}} \) and \( v_{\text{R}} \) are two sampling points on each side of the zero-crossing, \( f(v_{\text{L}}) \) and \( f(v_{\text{R}}) \) are the values of the two sampling points with opposite signs. Assuming the noise $z$ has a normal distribution with zero mean (\( z \sim N(0,\, \alpha_0) \)), the error of \( \mu \) is then calculated according to the propagation of errors:
\begin{equation}
	\mu^{\mathrm{e}} = \frac{f(v_{\text{L}})\cdot(v_{\text{R}}-v_{\text{L}})}{f(v_{\text{L}})-f(v_{\text{R}})}\cdot\sqrt{\left(\frac{\alpha_n}{f(v_{\text{L}})}\right)^{2}+\left(\frac{\sqrt{2}\alpha_n}{f(v_{\text{R}})}\right)^{2}}.
\end{equation}
As mentioned at the end of previous section, the errors of line-centers determined from the 1st- and 3rd-order derivatives are calculated (i.e., $n=1$ and $3$).
If two Gaussians are separated by an interval smaller than $\mu^{\mathrm{e}}$, they cannot be resolved. In other words, if they originate from adjacent LoSs, they are likely from the same emission structure. This concept forms the basis of the subsequent fitting and clustering processes.

\subsection{Multi-Gaussian spectral fitting with spatial-continuity constraints}
\label{sec::fitting}
Assuming that an emission structure does not vary significantly locally, the Gaussian parameters of adjacent LoSs should exhibit similar values. In the presence of noise, the fitted multi-Gaussian parameters for a single spectrum can be highly degenerate, with multiple parameter sets producing comparably good fits.
Additional information from spatially adjacent positions can reduce the multiplicity of solutions. For a Gaussian in the central LoS and one of its neighbor, with parameters $\{A_c,\,\mu_c,\,\sigma_c\}$ and $\{A_o,\,\mu_o,\,\sigma_o\}$, respectively, we define their relative line-center separation as
\begin{equation}
	\mu_{\mathrm{sep}}=\frac{|\mu_c-\mu_o|}{r_{\mathrm{v\_diff}}\cdot\min(\sigma_c,\,\sigma_o)+\mu^{\mathrm{e}}_{c}}
\end{equation}
where $\mu^{\mathrm{e}}_{c}$ is the error of line-center in the central LoS, and $r_{\mathrm{v\_diff}}$ is a manually defined hyper-parameter (typically $r_{\mathrm{v\_diff}}=1.22$ suggested by Rayleigh criterion).
Then we calculate the similarity between these two Gaussians as:
\begin{equation}
	s = 
	\begin{cases}
		1-\mu_{\mathrm{sep}}^{2},&\,\mu_{\mathrm{sep}} \le 1,\\
		0,&\,\mu_{\mathrm{sep}} > 1,
	\end{cases}
	\label{equ::los_similarity}
\end{equation}
which means that the two Gaussians are more similar when they are closer together. 

To establish correspondence between components in the central LoS and those in the neighboring LoSs, we construct a similarity matrix from Equation \ref{equ::los_similarity} and solve a one-to-one assignment problem.
The objective is to maximize the sum of the selected similarities, ensuring a global optimum matching between the Gaussian parameters in the two LoSs.
For each Gaussian, we search all neighboring LoSs within an angular radius of twice the angular resolution and select those with similar velocities according to Equation \ref{equ::los_similarity}.
If the number of kinematically close neighbors is smaller than a manually defined threshold, $min\_comp$, the central component is likely due to noise and is therefore removed from the central LoS.
Noise may also weaken a component to the point where it becomes indistinguishable from the noise, leading to the absence of the central component.
In such cases, if the number of neighbors is greater than $min\_comp$, a Gaussian inferred from neighbors is added to the central LoS.
In addition to adjusting the number of Gaussian components, the set of kinematically close neighbors also provides upper and lower bounds for the central Gaussian parameters during the subsequent fitting process. Specifically, the Gaussian parameters of the central components are constrained within the range defined by the minimum and maximum values of the corresponding parameters of the kinematically close Gaussians. Since smoothing does not improve the fitting results, we fit the original spectra directly.

Since wider Gaussians decrease faster than narrower ones as differentiation progresses, a wide and shallow component may sometimes be omitted, especially when decomposing \HI spectra. To avoid this potential mistake, several weak and wide Gaussians, with a small upper bound for $A$ and a large lower bound for $\sigma$, can be manually inserted as priors before the fitting process. We recommend adding at least one Gaussian to every LoS of pervasive \HI emission, and no additional components for CO data.

\subsection{Clustering with uncertainty and spatial-continuity constraints}
As a working hypothesis, we assume that each gas structure contains at most one Gaussian component per LoS; thus, different components within the same LoS cannot be assigned to the same structure. Starting with the LoS containing the strongest Gaussian component, every Gaussian in this LoS is regarded as the initial component of a pending structure.
Then neighboring LoSs are added, where each Gaussian is preferentially assigned to existing clusters by maximizing the total similarity $s$ between the current LoS and all existing structures using Equation \ref{equ::los_similarity}. If a Gaussian does not belong to any existing structure (i. e., $s$=0), a new pending structure is created, with this Gaussian as the initial component. A pending structure becomes a confirmed structure when the number of its LoSs exceeds $min\_comp$. 
This clustering process continues until all Gaussians, except for those with peak intensities lower than the noise level, are assigned to confirmed or pending structures.
Due to the fitting errors, a single structure may be split into several confirmed (or pending) structures, which often exhibit similar velocities at edge LoSs.
To address this, a merging step is performed: structures with continuous adjacent edges are combined, and pending structures are decomposed into components, which are then added to confirmed structures where possible.
The merging process continues until no further modifications can be made. Finally, all confirmed structures are regarded as real gas structures, while structures with their number of LoSs fewer than $min\_comp$ and those with peak intensities lower than the noise level are discarded.

To sum up, all hyper-parameters are list as follows:
\begin{itemize}
	\item $\sigma_{\text{min}}$: the minimum typical standard deviation line-width of the tracer that defined in Section \ref{sec::init_guess}. The default is the spectrum resolution.
	\item SNR$n$: the noise level for the $n$-th derivative that used in Equation \ref{equ::even_deri_constraints_noise}. The default is 3.
	\item $r_{\mathrm{v\_diff}}$: the ratio of line-center separation of two Gaussians to the narrower line-width among them that used in Equation \ref{equ::los_similarity}. The default is 1.22 (Rayleigh criterion).
	\item $min\_comp$: the minimum component number for a structure to be seen as a real structure (Section \ref{sec::fitting}). The default is 6.
\end{itemize}

\section{Application to Observations and Comparison}
\label{sec::application}
To test the general applicability of the proposed algorithm, we apply it to three different datasets: the \HI $4\pi$ survey (HI4PI), the Commensal Radio Astronomy FAST Survey (CRAFTS), and the Milky Way Imaging Scroll Painting (MWISP) survey.

\subsection{HI4PI}
The HI4PI all-sky data\footnote{\url{http://cdsarc.u-strasbg.fr/ftp/J/A+A/594/A116/}}\citep{2016A&A...594A.116H} were constructed by combining data from the Effelsberg-Bonn \HI Survey (EBHIS, \citealt{2011AN....332..637K,2016A&A...585A..41W}) and the Galactic All-Sky Survey (GASS, \citealt{2009ApJS..181..398M, 2010A&A...521A..17K}), with an angular resolution of \ang[angle-symbol-over-decimal]{;16.2} and spectral resolution of \SI{1.49}{\kms}. We use the prepared data on the Mollweide projection in equatorial coordinates. The hyper-parameters for decomposing the HI4PI data are listed in the first column of Table \ref{tab::hyperparameters}.

\begin{deluxetable*}{cccccc}
\tablenumm{1}
\tablecaption{The hyper-parameters for decomposing different datasets.\label{tab::hyperparameters}}
\tablewidth{0pt}
\tablehead{
\colhead{} & \multirow{2}{*}{HI4PI} & \multirow{2}{*}{CRAFTS} & \colhead{MWISP} & \colhead{MWISP} & \colhead{MWISP}\\
\cline{4-6}
\colhead{} & \colhead{} & \colhead{} & \colhead{$^{12}\text{CO}$} & \colhead{$^{13}\text{CO}$} & \colhead{$\text{C}^{18}\text{O}$}
}
\startdata
$\sigma_{\text{min}}$ ($\text{km\,s}^{-1}$) & 1.5 & 1.5 & 1.0 & 1.0 & 1.0\\
SNR0 & 9 & 9 & 9 & 9 & 9\\
SNR2 & 2 & 2 & 2 & 2 & 2\\
SNR4 & 1 & 1 & 1 & 1 & 1\\
$r_{\mathrm{v\_diff}}$ & 1.22 & 1.22 & 1.22 & 1.22 & 1.22\\
$min\_comp$ & 6 & 6 & 6 & 6 & 6\\
\enddata
\end{deluxetable*}

\subsubsection{Component-wise Results}
\begin{figure*}[!ht]
	\plotone{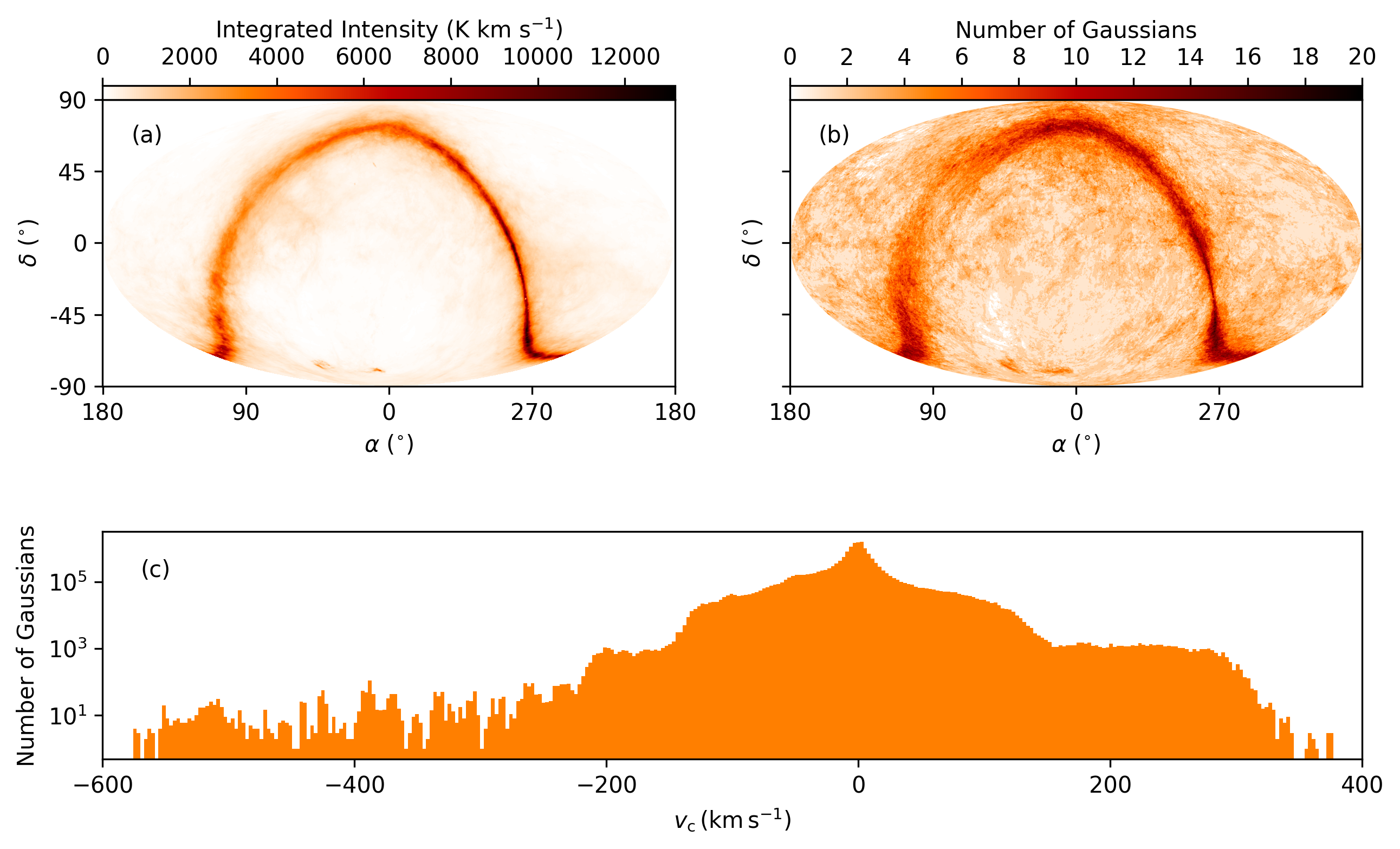}
	\caption{(a) Integrated intensity map of HI4PI data, (b) distribution of the number of Gaussians in equatorial coordinates, and (c) histogram of central velocities of Gaussians. The color scales in (a) and (b) represent the integrated intensity and the number of Gaussians, respectively.}
	\label{fig::m0_and_comp_num_in_lb_HI4PI_mol}
\end{figure*}

The term ``component" here denotes a Gaussian component derived from the multi-Gaussian fitting process, with those discarded by the clustering process already removed.
A total of \num[group-separator={,}]{16788310} Gaussian components (after discarding structures with fewer than $min\_comp=6$ components) were extracted from the HI4PI all-sky data, corresponding to an average of $\sim$2.8 Gaussians per LoS. Figure \ref{fig::m0_and_comp_num_in_lb_HI4PI_mol} presents the integrated intensity map of the original HI4PI data, along with the distribution of the number of Gaussians along each LoS in equatorial coordinates, and the histogram of their velocity distribution.
Overall, the spatial distribution of Gaussians closely follows the \HI integrated intensity map, with approximately 20 velocity components on the Galactic plane. As shown in Figure \ref{fig::m0_and_comp_num_in_lb_HI4PI_mol}(c), most Gaussians concentrate around \SI{0}{\kms}, with additional prominent sub-structures appearing at $\sim\pm$\SI{120}{\kms} and $\sim$\SI{-200}{\kms}.

\begin{figure*}
	\gridline{\fig{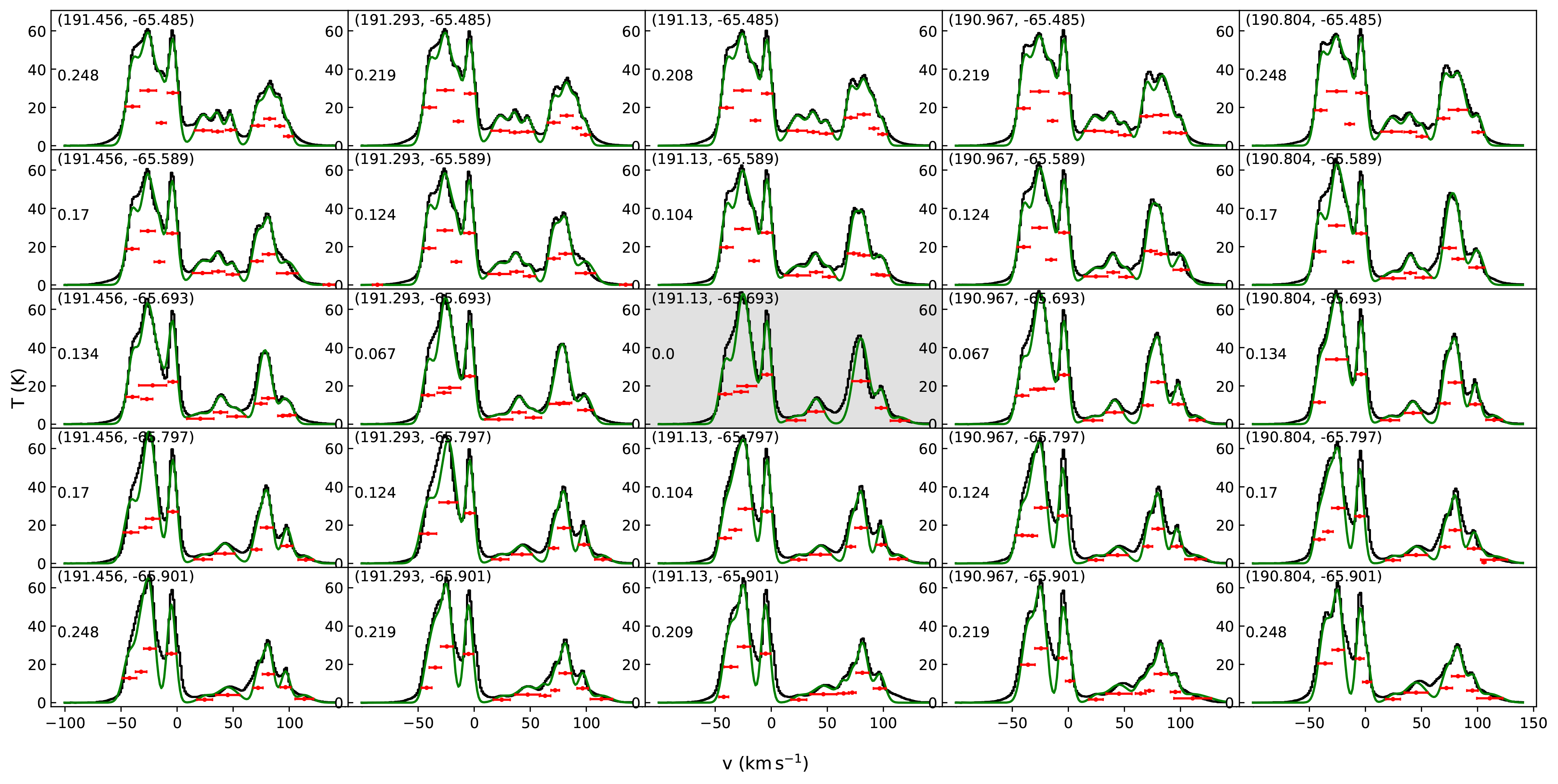}{0.95\textwidth}{(a)}}
	\gridline{\fig{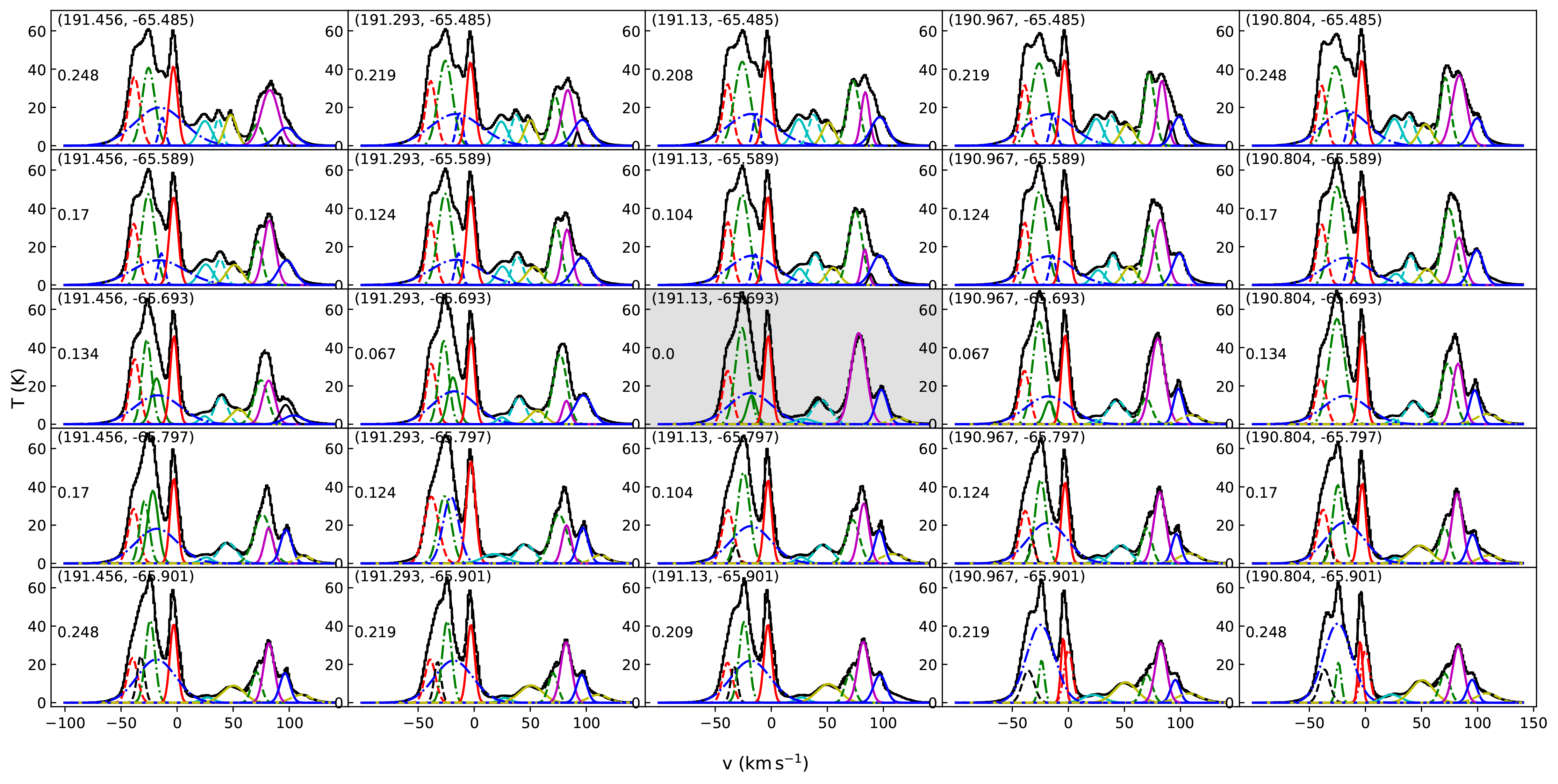}{0.95\textwidth}{(b)}}
	\caption{Initial estimates (a) and clustering results (b) for 25 spatially adjacent \HI spectra. In both panels, black solid lines represent the observed spectra. In (a), red dots and bars denote the estimated Gaussians, where the $x$-coordinate of a red dot corresponds to the line-center, the $y$-coordinate represents half of the peak intensity, and the bar width indicates the FWHM. Green solid lines show the total intensity of these Gaussians. In (b), colored lines depict the multi-Gaussian fitting results, with lines of the same color and style indicating Gaussians grouped into the same structure. The black dashed line represents the total intensity of all Gaussians. The numbers in the top-left parentheses of each LoS indicate the ($\alpha,\,\delta$) coordinate values, while the middle-left numbers denote the angular separation from the central LoS.}
	\label{fig::HI4PI_mol_clus_guess}
\end{figure*}

Figure \ref{fig::HI4PI_mol_clus_guess} presents the initial guesses of Gaussian parameters and the clustering results (after fitting) for 25 spatially adjacent \HI spectra.
Although some degree of multiplicity remains in the fitting process, the overall fitting and clustering results are considered reliable.

\begin{figure}[!ht]
	\plotone{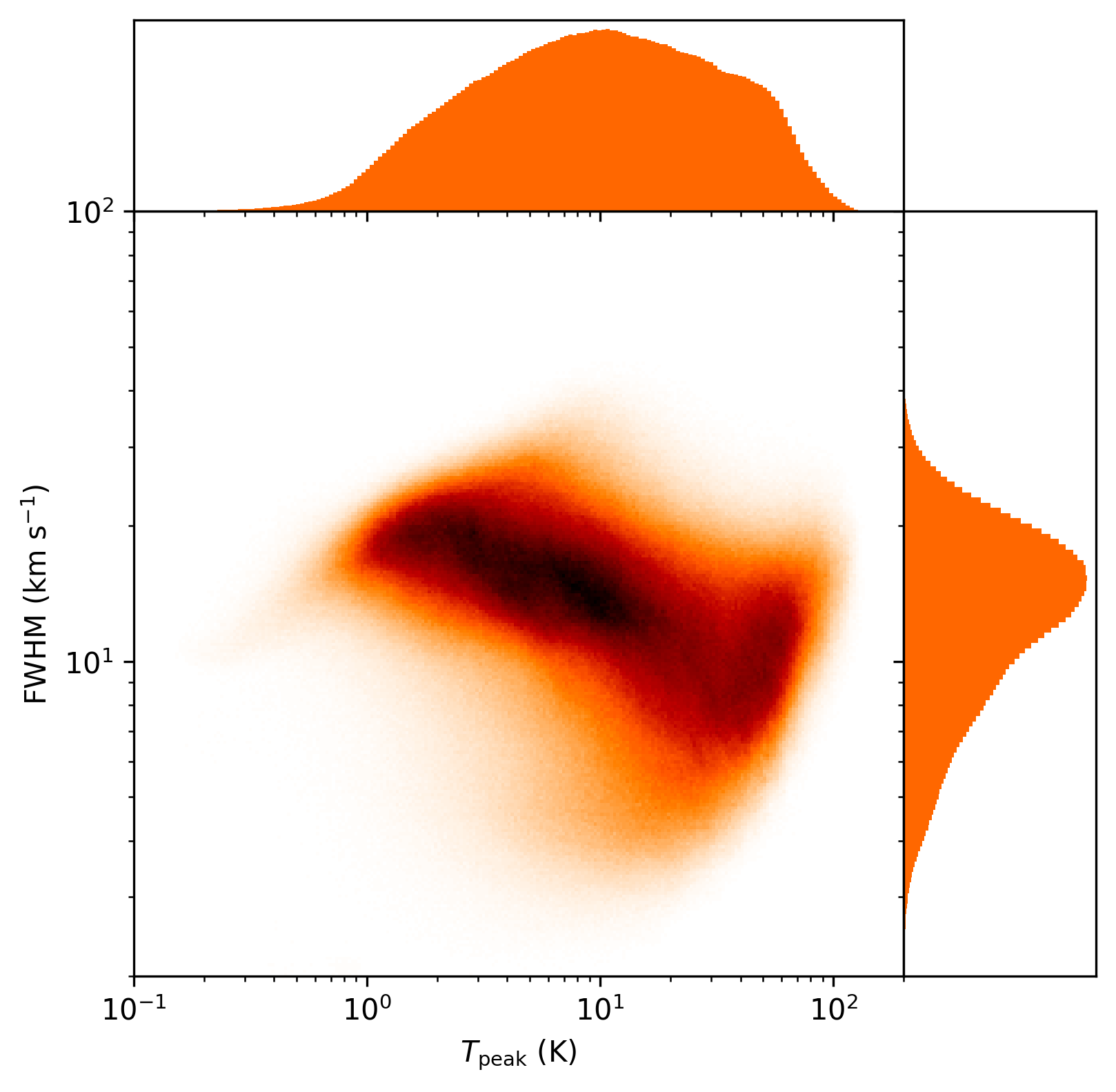}
	\caption{Density map of Gaussian fit parameters ($T_{\mathrm{peak}}$, FWHM) for all HI4PI components, where darker regions indicate higher densities in parameter space. The top and right panels display the marginal histograms of $T_{\mathrm{peak}}$ and FWHM, respectively.}
	\label{fig::comp_wise_T_w_relation_HI4PI_mol}
\end{figure}

Figure \ref{fig::comp_wise_T_w_relation_HI4PI_mol} illustrates the correlation between the peak intensity ($T_{\mathrm{peak}}$) and FWHM for all Gaussian components, along with their respective histograms.
The density map reveals two distinct regimes: the fainter but broader Gaussians (belonging to WNM or UNM according to their line-widths) imply a negative correlation between the FWHM and intensity, while the FWHM of brighter Gaussians (belonging to CNM or UNM) is positively correlated with their intensities.

It is important to note that due to the presence of \HI self-absorption (HISA) features, fitting the spectra with a multi-Gaussian emission model can mistakenly break the WNM down into multiple narrower Gaussians. This process leads to an underestimation of the warm gas fraction and overestimation of the other two phases.

\subsubsection{Structure-wise Results}
\begin{figure*}[!ht]
	\includegraphics[width=1.\textwidth]{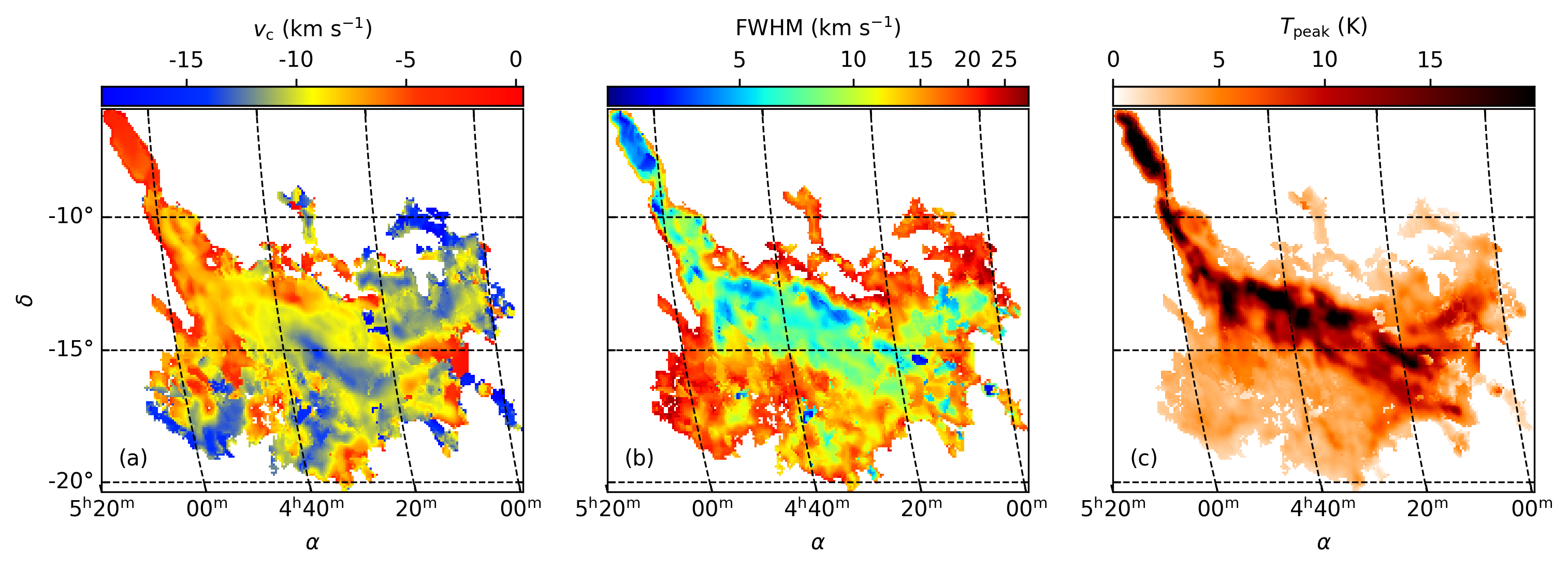}
	\caption{The (a) line-center velocity, (b) FWHM line-width, and (c) peak intensity map of one \HI structure in the HI4PI all-sky decomposition result.}
	\label{fig::largest_HI4PI_mol}
\end{figure*}

The term ``structure" here denotes an ensemble of neighboring Gaussian components with similar line-center velocities.
The all-sky HI4PI data is decomposed into \num[group-separator={,}]{45299} structures.
Figure \ref{fig::largest_HI4PI_mol} presents one example of \HI structure obtained from GDCluster.
This structure spans from \SI{-67}{\kms} to \SI{-31}{\kms}, with a mean FWHM of \SI{15.92}{\kms} and mean peak intensity of \SI{2.4}{K}.

\begin{figure}[!ht]
	\plotone{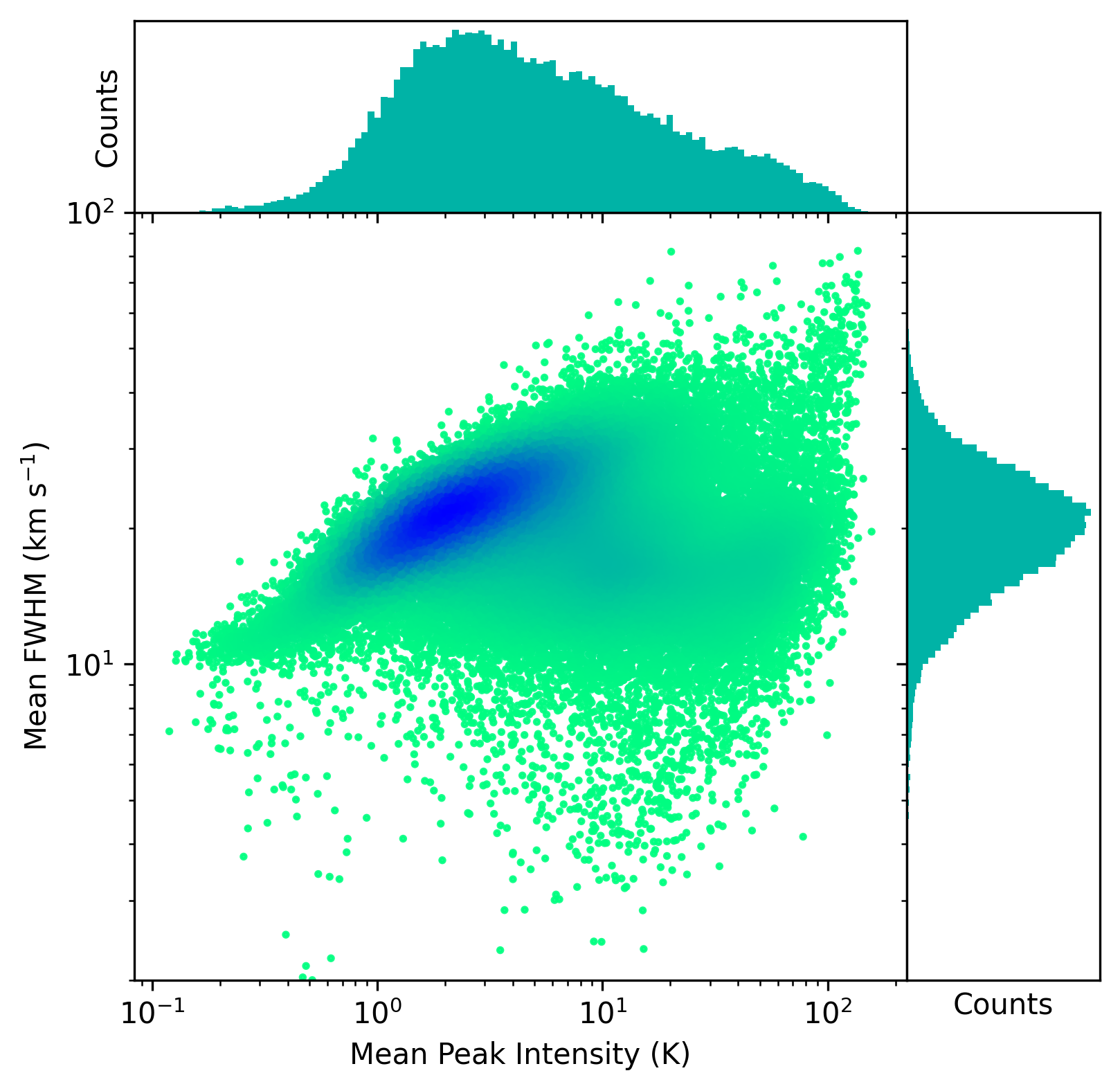}
	\caption{Density map of mean peak intensity versus FWHM for all HI4PI \HI structures. Each point represents an individual structure, with the blue indicating denser regions in parameter space. The top and right panels display the marginal histograms of mean peak intensities and FWHM, respectively.}
	\label{fig::struc_wise_inten_w_relation_HI4PI_mol}
\end{figure}

As an example of statistical results in decomposition structures, we show the relationship between mean peak intensities and FWHM of all HI4PI \HI structures in Figure \ref{fig::struc_wise_inten_w_relation_HI4PI_mol}.
Most structures exhibit a mean peak intensity of approximately \SI{2}{K} and an FWHM of around \SI{20}{\kms}. The overall distribution can be categorized into two primary types:
structures in the fainter and wider region, where the mean FWHM gradually increases with peak intensity, and the rest with smaller FWHM but higher peak intensities, which are more dispersed in the parameter space.
However, we cannot directly associate these two types of structures with CNM or WNM, since the histogram of the mean FWHM in the right panel of Figure \ref{fig::struc_wise_inten_w_relation_HI4PI_mol} shows no distinct separation. A single \HI structure may contain CNM, UNM, and WNM components, and can smoothly transition from one phase to another (see Figure \ref{fig::largest_HI4PI_mol}).

\subsection{CRAFTS}

The CRAFTS survey is conducted using the Five Hundred Meter Aperture Spherical Telescope (FAST) and simultaneously carries out multiple surveys in a single drift-scan \citep{2018IMMag..19..112L}.
The CRAFTS \HI Narrow All-Sky Survey aims to map the Galactic \HI emission with a beam-width of \ang{;3} (regridded into a pixel size of \ang{;1} $\times$ \ang{;1}) and a velocity resolution of \SI{0.1}{\kms}.
The data used in this analysis\footnote{\url{https://hiverse.zero2x.org}} cover the sky region of $90^\circ \le \alpha \le 100^\circ$ and $-13^\circ \le \delta \le 3^\circ$ (\ang{10} $\times$ \ang{10}).
After manual inspection, we select data in the velocity range of \SI{-40}{\kms} $\le v \le$ \SI{110}{\kms}, where the main emission resides, to implement the decomposition. The hyper-parameters used for decomposing the CRAFTS data are listed in the second column of Table \ref{tab::hyperparameters}.

\begin{figure*}[!ht]
	\fig{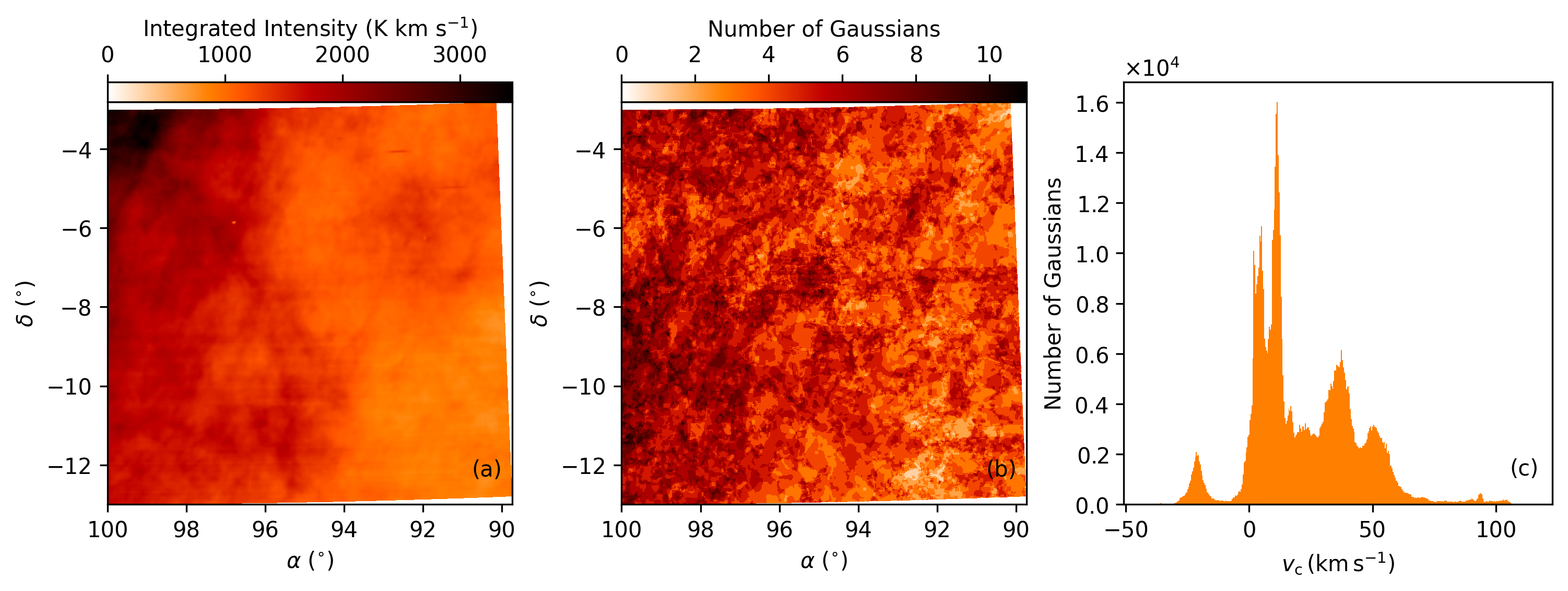}{0.8\textwidth}{}
	\caption{The same as Figure \ref{fig::m0_and_comp_num_in_lb_HI4PI_mol} for CRAFTS \HI data in a \ang{10} $\times$ \ang{10} sub-region.}
	\label{fig::m0_and_comp_num_in_lb_FAST}
\end{figure*}

In total, \num[group-separator={,}]{766623} Gaussians are extracted in this region, with 4.791 Gaussians per LoS.
The distribution of these Gaussians and the integrated intensity map are shown in Figure \ref{fig::m0_and_comp_num_in_lb_FAST}.
The number of Gaussians per LoS does not align well with the distribution of total flux in this region. The velocity distribution reveals a few distinct regimes.

\begin{figure}[!ht]
	\plotone{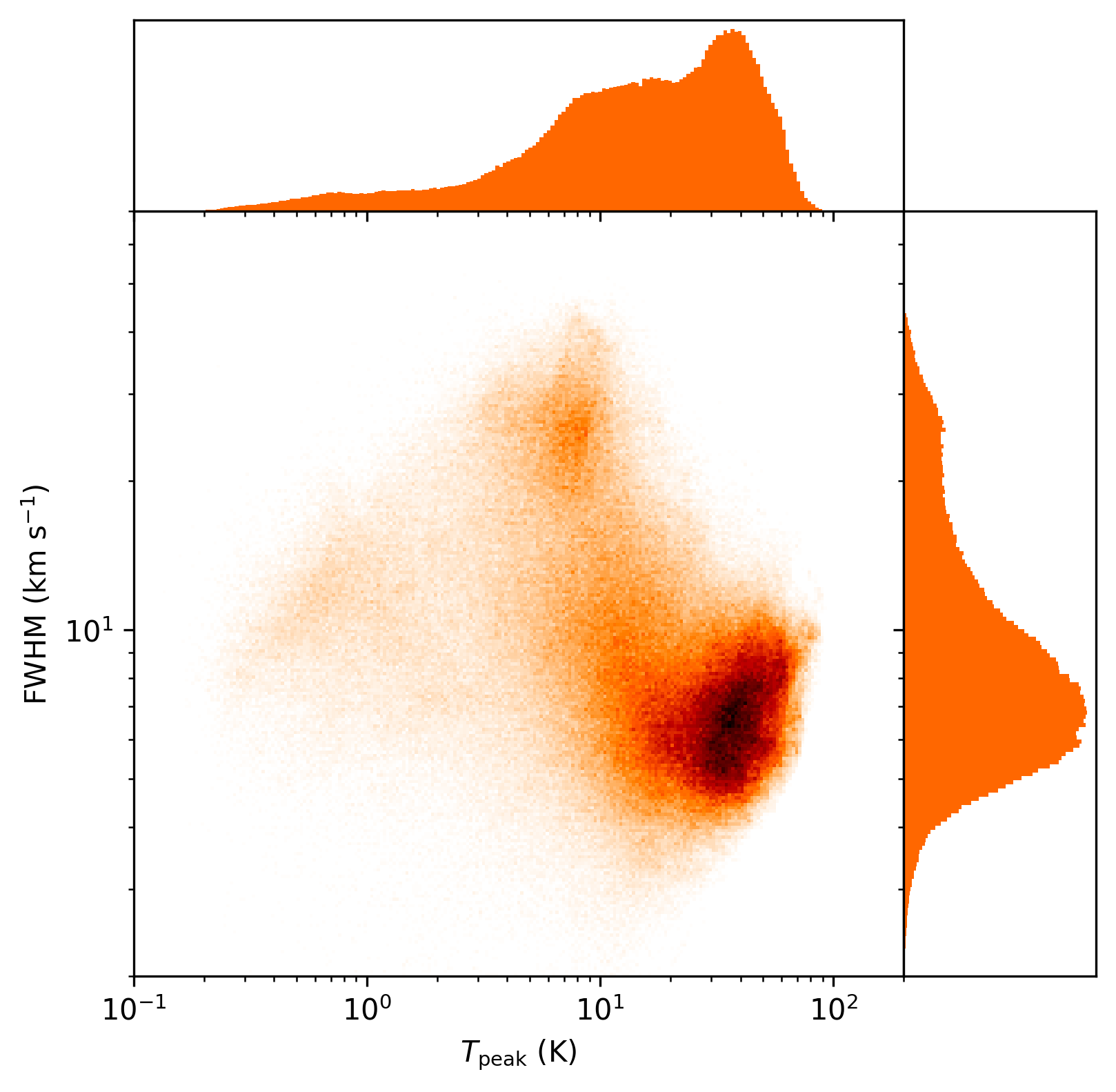}
	\caption{The same as Figure \ref{fig::comp_wise_T_w_relation_HI4PI_mol} for CRAFTS \HI Gaussians.} \label{fig::comp_wise_T_w_relation_FAST}
\end{figure}

\begin{figure}[!ht]
	\plotone{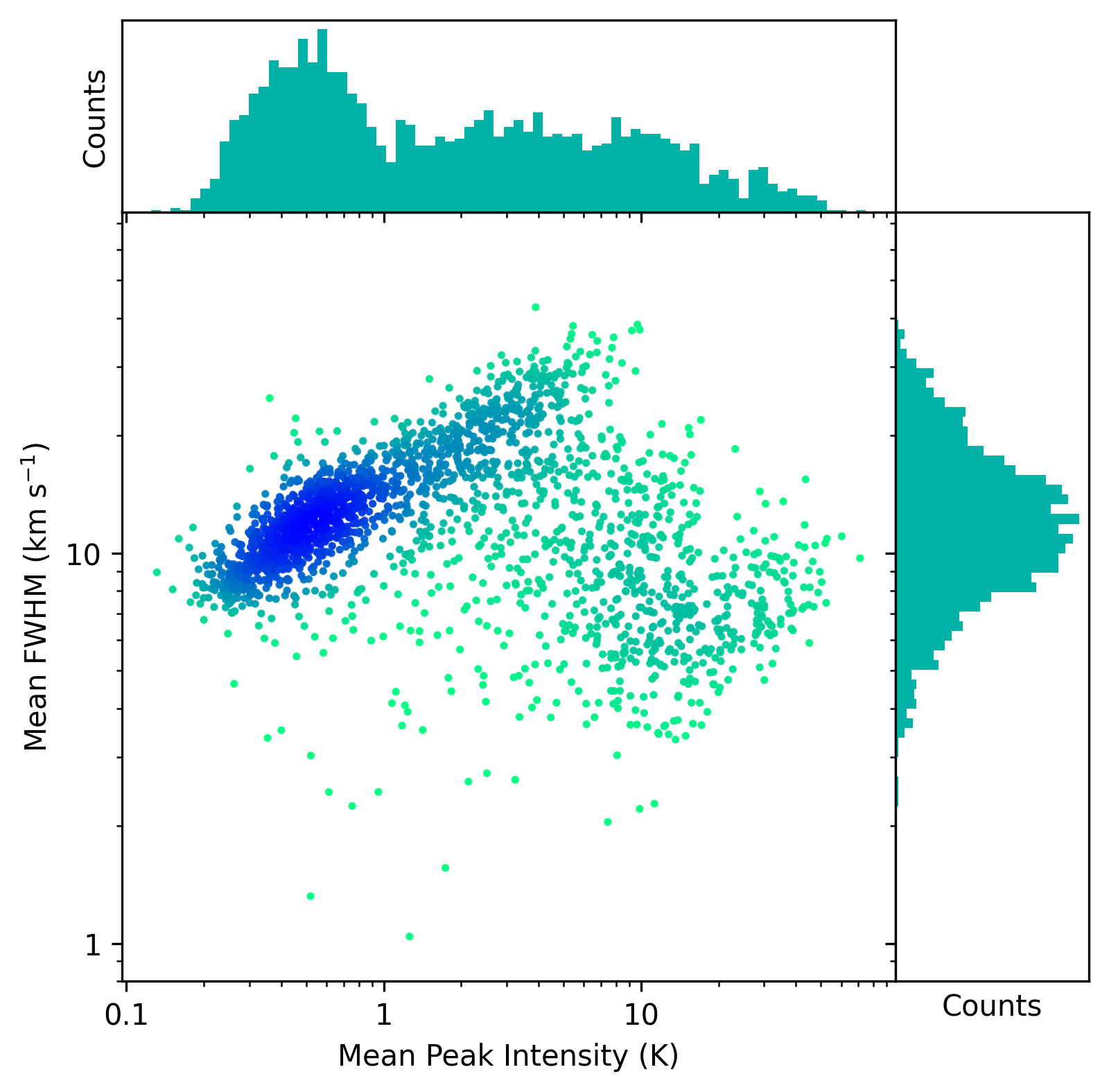}
	\caption{The same as Figure \ref{fig::struc_wise_inten_w_relation_HI4PI_mol} for CRAFTS \HI structures.}
	\label{fig::struc_wise_inten_w_relation_FAST}
\end{figure}

The correlation between $T_{\mathrm{peak}}$ and FWHM for all Gaussians is shown in Figure \ref{fig::comp_wise_T_w_relation_FAST}, where we can also see distinct \HI phases, similar to those in the HI4PI Gaussians.
It is the same situation as the structure-wise statistics shown in Figure \ref{fig::struc_wise_inten_w_relation_FAST} for \num[group-separator={,}]{2247} structures, where two different types of regimes can be seen.

\subsection{MWISP}
\label{sec::MWISP_decomp}
The MWISP\footnote{\url{http://english.dlh.pmo.cas.cn/ic/in/}} project \citep{2019ApJS..240....9S}, which is conducted using PMO-\SI{13.7}{m} millimeter-wavelength telescope at Delingha, China, observes $\ce{^{12}CO}$, $\ce{^{13}CO}$, and $\ce{C^{18}O}$ ($J$=1--0) transitions simultaneously. This project covers the region of \ang[angle-symbol-over-decimal]{9.75} $\le l\le$ \ang[angle-symbol-over-decimal]{229.75} and \ang[angle-symbol-over-decimal]{-5.25} $\le b\le$ \ang[angle-symbol-over-decimal]{5.25}, and we use a sub-region (\ang[angle-symbol-over-decimal]{104.75} $\le l\le$ \ang[angle-symbol-over-decimal]{150.25}, \ang[angle-symbol-over-decimal]{-5.25} $\le b\le$ \ang[angle-symbol-over-decimal]{5.25}) in the second quadrant to validate our algorithm. The data have a velocity resolution of \SI{0.16}{\kms}, \SI{0.17}{\kms}, and \SI{0.17}{\kms} for $\ce{^{12}CO}$, $\ce{^{13}CO}$, and $\ce{C^{18}O}$, respectively, and a cell size of \ang{;;30}~$\times$~\ang{;;30} for all three tracers.

\begin{figure*}[ht!]
    \centering
    \includegraphics[width=0.7\textwidth]{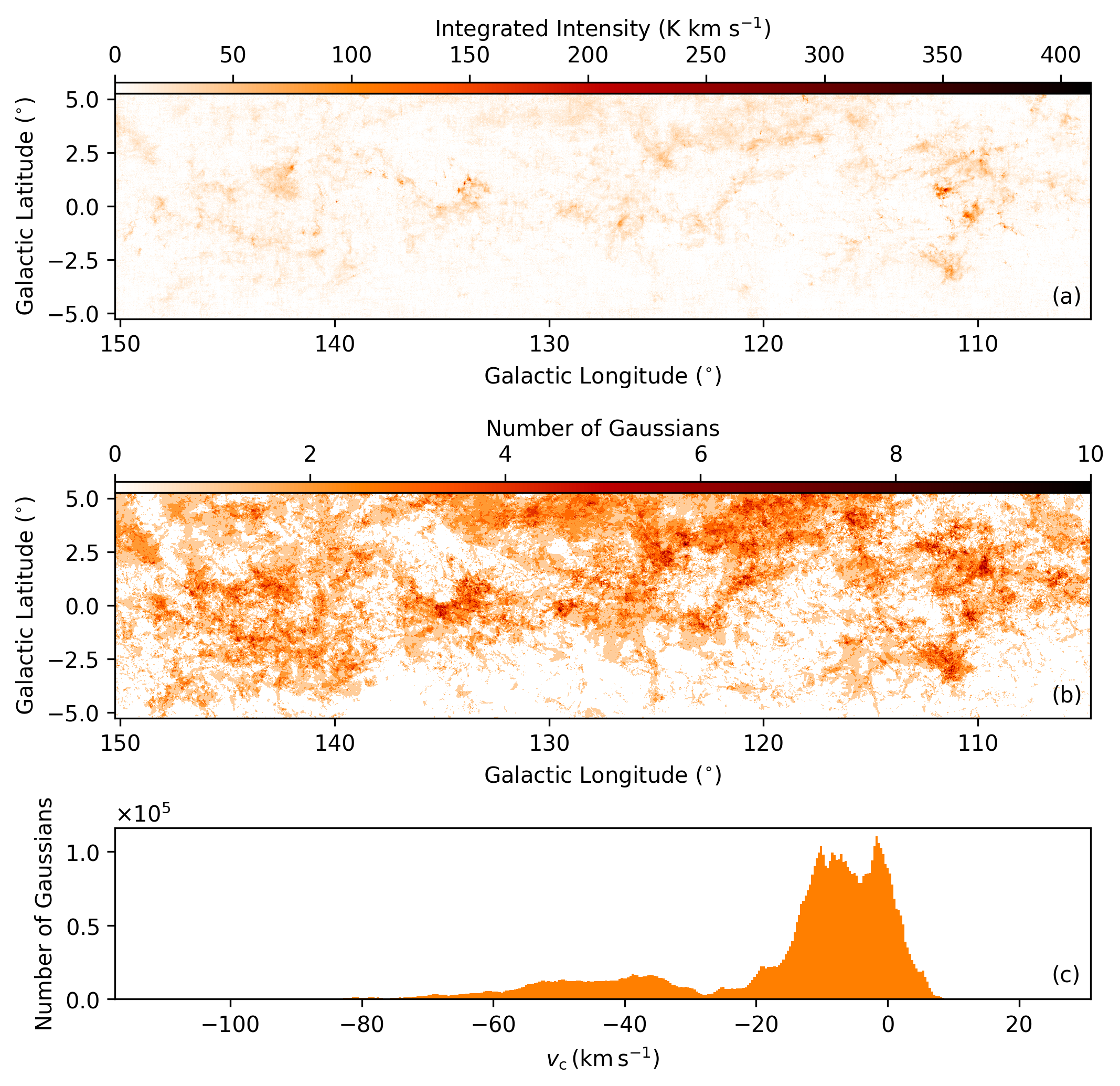}
    \caption{The same as Figure \ref{fig::m0_and_comp_num_in_lb_HI4PI_mol} for $\ce{^{12}CO}$ data. \label{fig::m0_and_comp_num_in_lb_12CO}}
\end{figure*}

In total, \num[group-separator={,}]{6886321}/\num[group-separator={,}]{1149185}/\num[group-separator={,}]{14483} Gaussian components (on average 0.912/0.167/0.002 Gaussians per LoS) are extracted in the $\ce{^{12}CO}$/$\ce{^{13}CO}$/$\ce{C^{18}O}$ datacubes. Hereafter in the main text we only present the results of $\ce{^{12}CO}$, and the $\ce{^{13}CO}$/$\ce{C^{18}O}$ decomposition results are shown in Appendix \ref{apx::stats_13CO_C18O}. The integrated intensity of the raw datacube and distribution of $\ce{^{12}CO}$ Gaussians are shown in Figure \ref{fig::m0_and_comp_num_in_lb_12CO}.
Approximately 44.06\% of LoSs contain no Gaussian components, 8.13\% contain at least three, and 0.34\% contain more than five. The number of components generally correlates with the integrated intensity along the LoS, except in certain star-forming regions, such as the Cep OB3 ($l\sim$ \ang{110}, $b\sim$ \ang{2}), where the velocity structure is significantly more complex than in other regions. Figure \ref{fig::m0_and_comp_num_in_lb_12CO}(c) presents the velocity distribution of all components, which are predominantly concentrated in the Local and Perseus Arms.

\begin{figure}[ht!]
    \plotone{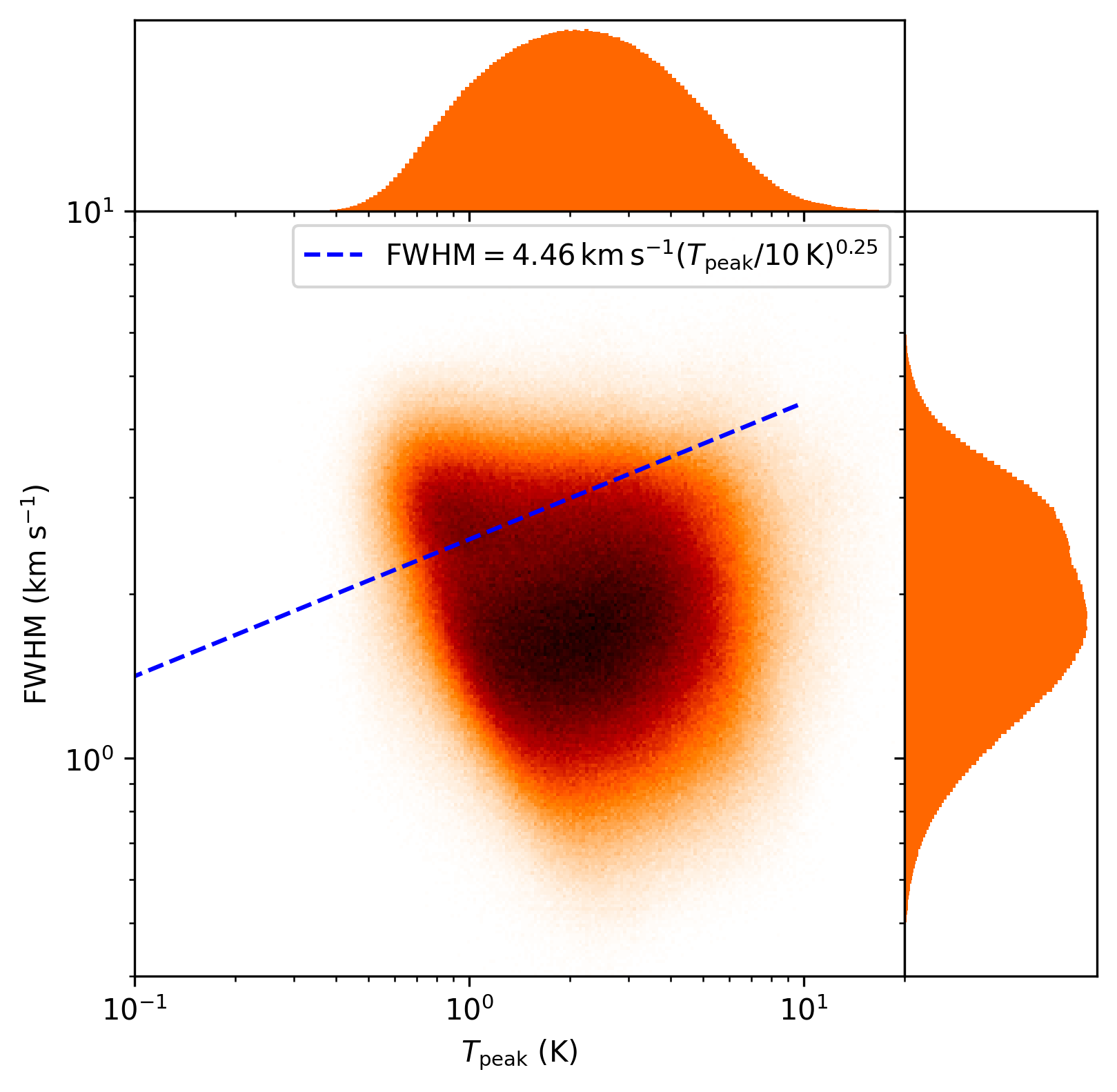}
    \caption{The same as Figure \ref{fig::comp_wise_T_w_relation_HI4PI_mol} for $^{12}\mathrm{CO}$ Gaussians. The blue dashed line denotes the bound of two distinct clusters.}
    \label{fig::comp_wise_T_lw}
\end{figure}

Figure \ref{fig::comp_wise_T_lw} shows the correlation between $T_{\mathrm{peak}}$ and FWHM for all Gaussians.
The distribution of $T_{\mathrm{peak}}$ is approximately log-normal, while the FWHM is likely distributed in two log-normals that peak at \SI{1.7}{\kms} and \SI{3.0}{\kms}. The joint distribution clearly reflects this issue: components diverge into two clusters, one with a larger FWHM value but lower peak intensity, another with a smaller FWHM value but higher peak intensity. These two types of Gaussians can be divided by a function
$$\mathrm{FWHM}=\SI{4.46}{\kms}\left(\frac{T_{\mathrm{peak}}}{\SI{10}{K}}\right)^{0.25},$$
corresponding to the blue dashed line in Figure \ref{fig::comp_wise_T_lw}.
We note that the coefficient and exponent remain unchanged under variations of any hyper-parameter, indicating the intrinsic property of molecular gas.

\begin{figure}[ht!]
    \plotone{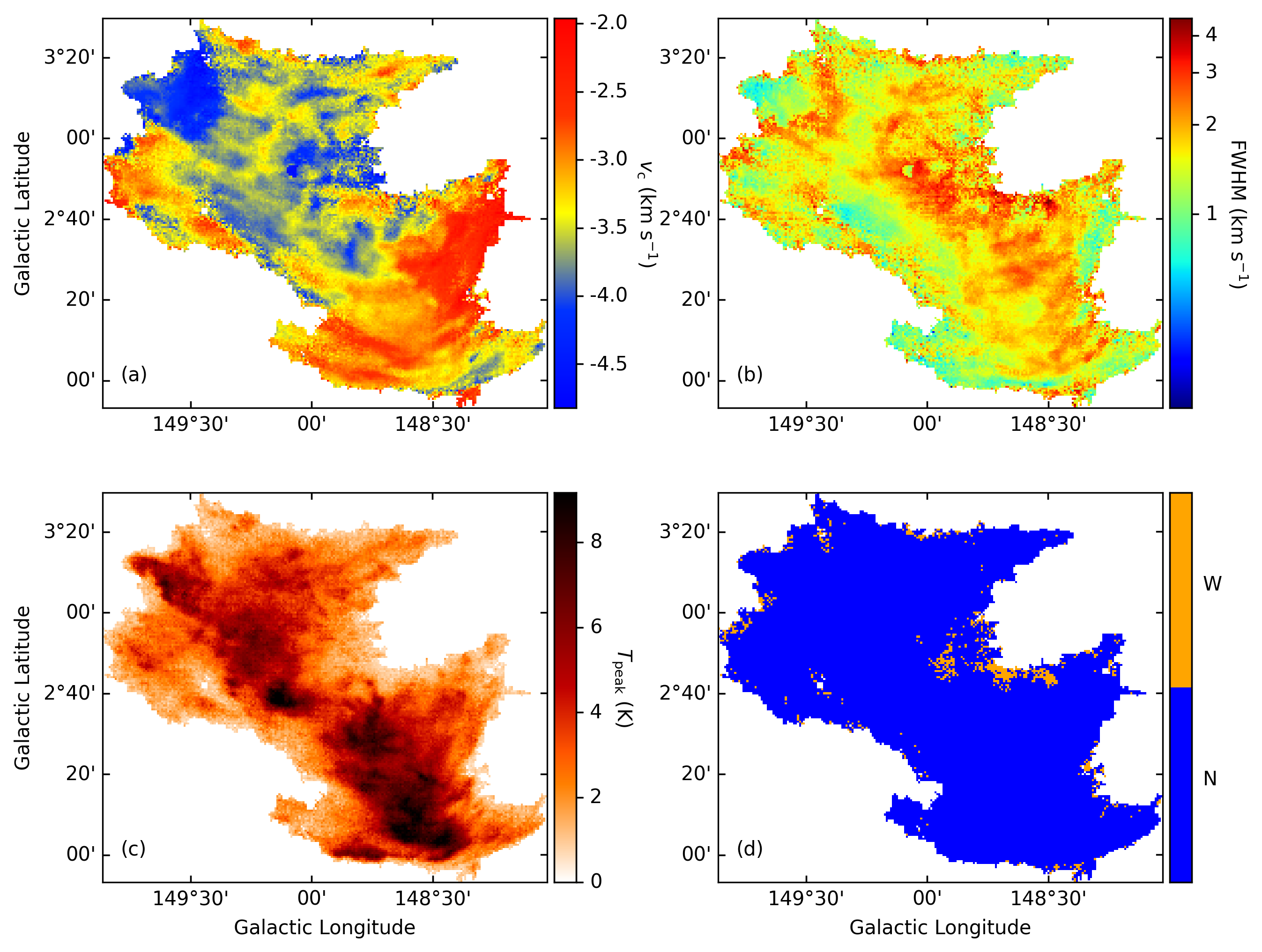}
    \caption{The same as Figure \ref{fig::largest_HI4PI_mol} for a $\ce{^{12}CO}$ structure. In panel (d), the wider (W) type of Gaussians in Figure \ref{fig::comp_wise_T_lw} are painted in yellow, while the narrower (N) type in blue. \label{fig::12CO_struc}}
\end{figure}

We identified \num[group-separator={,}]{47119}/\num[group-separator={,}]{12360}/\num[group-separator={,}]{395} $^{12}\text{CO}$/$^{13}\text{CO}$/$\text{C}^{18}\text{O}$ structures, respectively. Figure \ref{fig::12CO_struc} displays one $\ce{^{12}CO}$ structure identified by GDCluster.
In Figure \ref{fig::12CO_struc}(d) we note that the narrow type of Gaussians in Figure \ref{fig::comp_wise_T_lw} is more concentrated in its distribution, while the wide type is more dispersed.
However, the physical origin of this distinction remains unclear.

\begin{figure*}[ht!]
    \plotone{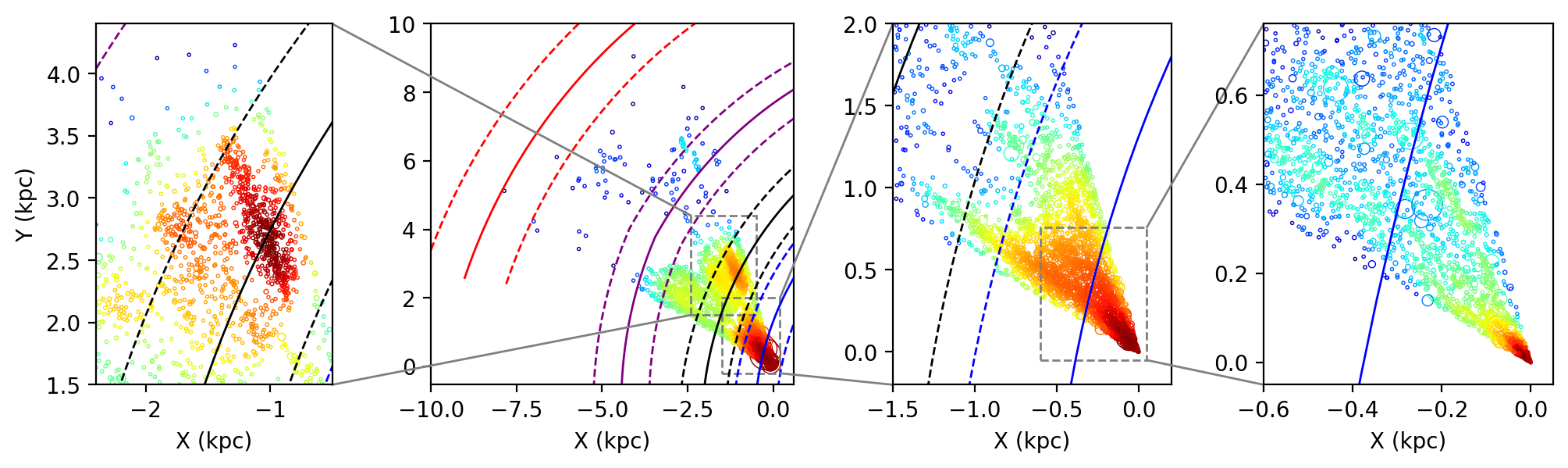}
    \caption{Face-on view of $\ce{^{12}CO}$ structures with an angular size larger than \SI{30}{arcmin^{2}}. Circles represent the structures, with their radius indicating relative sizes and the color representing the densities of structures in space. The Sun is located at (0, 0)\,kpc, and the Galactic center is at (8.15, 0)\,kpc. The first and third plots show sub-regions of the second one, as indicated by the gray dashed lines and boxes, while the fourth plot is a sub-region of the third one. Blue, black, purple, and red solid lines trace the centers of the Local, Perseus, Norma-Outer, and Outer Scutum–Centaurus arms, respectively, with dashed lines indicating their widths \citep{reid_trigonometric_2019}.\label{fig::struc_scatter_face_on_zoom}}
\end{figure*}

Taking the mean line-center velocity of all Gaussian components within a structure as its representative velocity, we calculate the distances of all $\ce{^{12}CO}$ structures using the Galactic rotation curve from \citet{reid_trigonometric_2014}. 
Additionally, we correct for non-circular (peculiar) motions of sources located within \SI{6}{kpc} of the Sun, following the method described in \citet{2023ApJS..268...46L}.
The face-on view of these structures is shown in Figure \ref{fig::struc_scatter_face_on_zoom}.
To better visualize the distribution of structure sizes in crowded regions, we masked out structures with angular areas smaller than \SI{30}{arcmin^{2}}.
While $\ce{^{12}CO}$ structures are predominantly concentrated along the spiral arms, a substantial number—including some large structures—are also found in the inter-arm regions.

\subsection{Comparison with other algorithms}
\label{sec::comparison}
\subsubsection{Molecular structure in the MWISP survey}
\label{sec::stats_compare}
\begin{figure*}[ht!]
    \centering
    \includegraphics[width=0.7\textwidth]{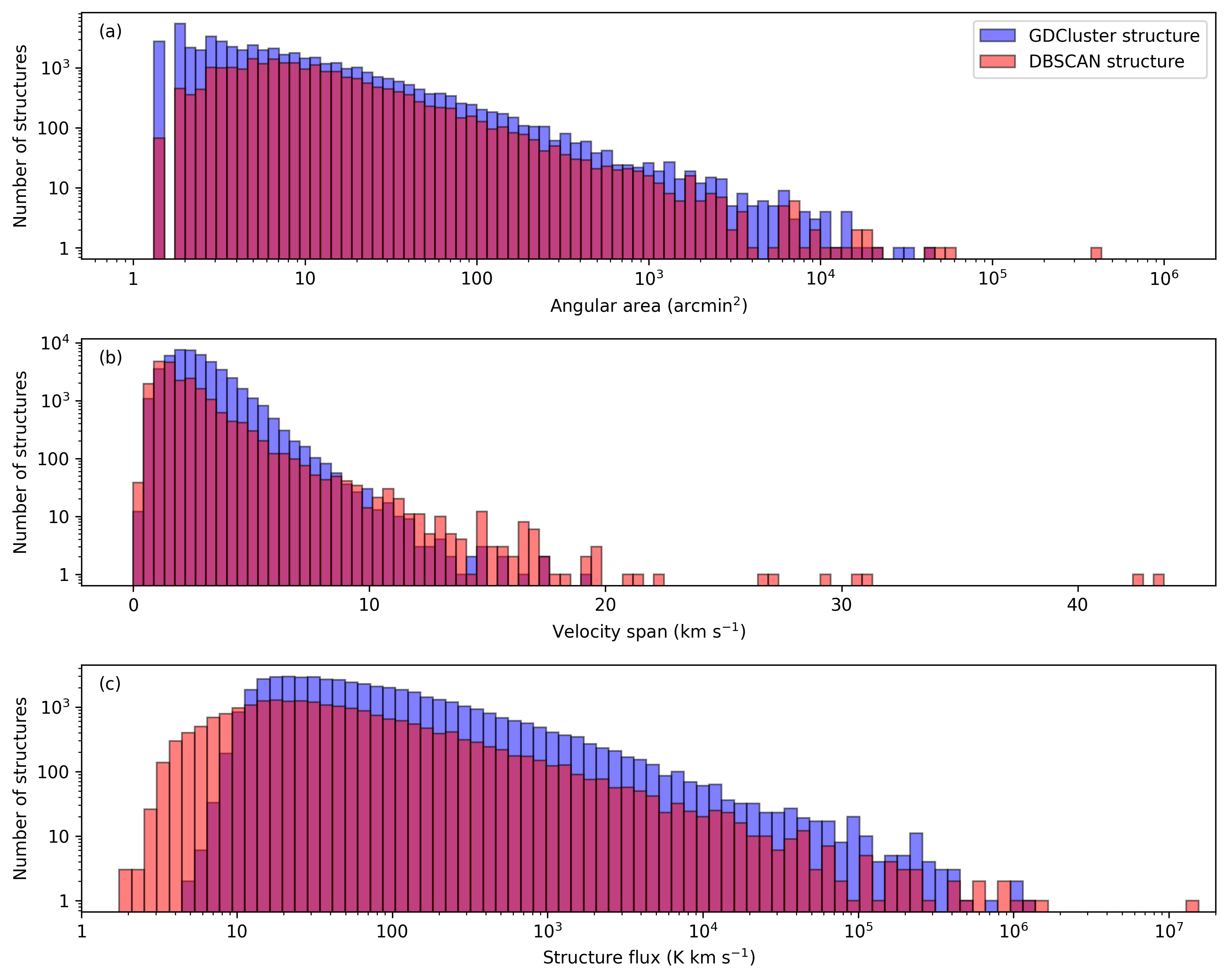}
    \caption{Histogram of (a) angular area, (b) velocity span, and (c) total flux of GDCluster (blue) and DBSCAN (red) structures.}
    \label{fig::dbscan_compare_hists}
\end{figure*}

Density-Based Spatial Clustering of Applications with Noise (DBSCAN) algorithm has been used to decompose MWISP survey data into molecular structures \citep{2020ApJ...898...80Y,2021A&A...645A.129Y,2021ApJS..257...51Y}.
With the DBSCAN algorithm already fine-tuned by these authors, we used the cloud catalog provided by \citet{2021A&A...645A.129Y} to compare structure decomposition results. In addition, we tuned and applied the GaussPy+ algorithm to the same region. Since GaussPy+ decomposes spectra only into Gaussian components rather than structures, it is included only in the spectral comparison.
In the \ang[angle-symbol-over-decimal]{104.75} $\le l\le$ \ang[angle-symbol-over-decimal]{150.25}, \ang[angle-symbol-over-decimal]{-5.25} $\le b\le$ \ang[angle-symbol-over-decimal]{5.25} region, \num[group-separator={,}]{21351} molecular structures are detected by DBSCAN, while \num[group-separator={,}]{47119} structures are detected by GDCluster in the same region. The histograms of angular area, velocity span, and total flux of structures detected by DBSCAN and GDCluster are shown in Figure \ref{fig::dbscan_compare_hists}.
The angular area distribution of GDCluster structures exhibits power-law behaviour across nearly the entire range, whereas for DBSCAN structures the same power-law holds only for those with angular areas larger than \SI{10}{arcmin^2}. Specifically, there are far fewer small-scale DBSCAN structures ($<\SI{7}{arcmin^2}$) but a few significantly larger ones ($>\SI{5e4}{arcmin^2}$).
The velocity span of GDCluster structures ranges from $\sim\SI{0.3}{\kms}$ to $\sim\SI{19.0}{\kms}$, whereas DBSCAN structures range from $\SI{0}{\kms}$ to $\sim\SI{43.7}{\kms}$. Figure \ref{fig::dbscan_compare_hists}(b) shows that the velocity span distribution of GDCluster structures is steeper than that of DBSCAN structures.
The largest DBSCAN structure ($\sim\SI{3.9e5}{arcmin^{2}}$, corresponding to Figure 5 and 6 in \citet{2021A&A...645A.129Y}) also has the widest velocity range ($\sim\SI{43.7}{\kms}$), resulting in a total flux ($\sim\SI{1.35e7}{K.\kms}$) nearly one order of magnitude greater than that of any other structure, as shown in Figure \ref{fig::dbscan_compare_hists}(c). Furthermore, the number of GDCluster structures exceeds that of DBSCAN structures in nearly every interval above $\sim\SI{10}{K.\kms}$, but is lower for fluxes below $\sim\SI{10}{K.\kms}$.

\begin{figure*}[ht!]
    \centering
    \includegraphics[width=0.7\textwidth]{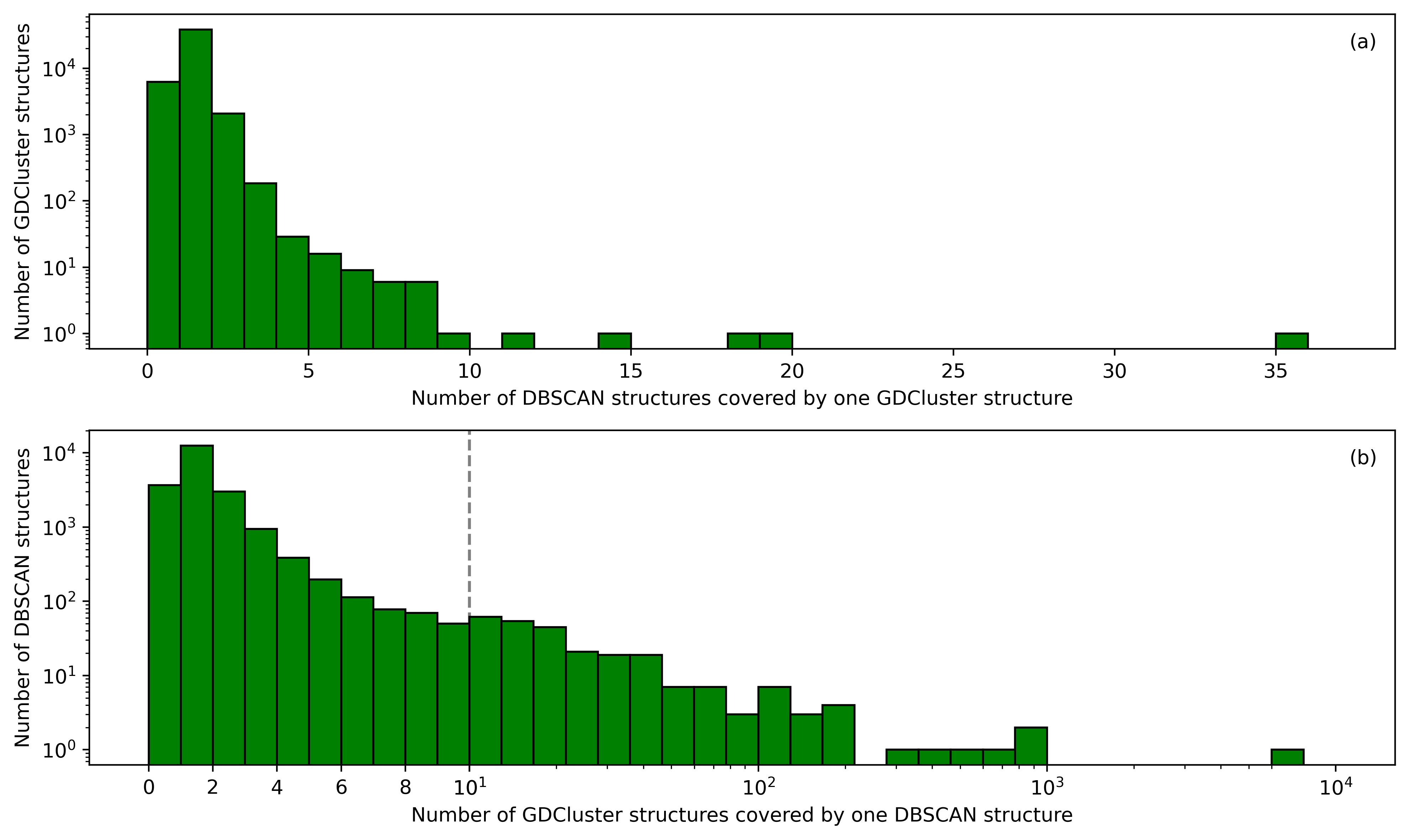}
    \caption{Histogram of (a) GDCluster structures covered by one DBSCAN structure and (b) DBSCAN structures covered by one GDCluster structure. The gray dashed line in panel (b) separates the $x$-axis into linear- and logarithmic-scale regions.}
    \label{fig::one_cover_multi}
\end{figure*}

\begin{figure*}[ht!]
    \centering
    \includegraphics[width=0.9\textwidth]{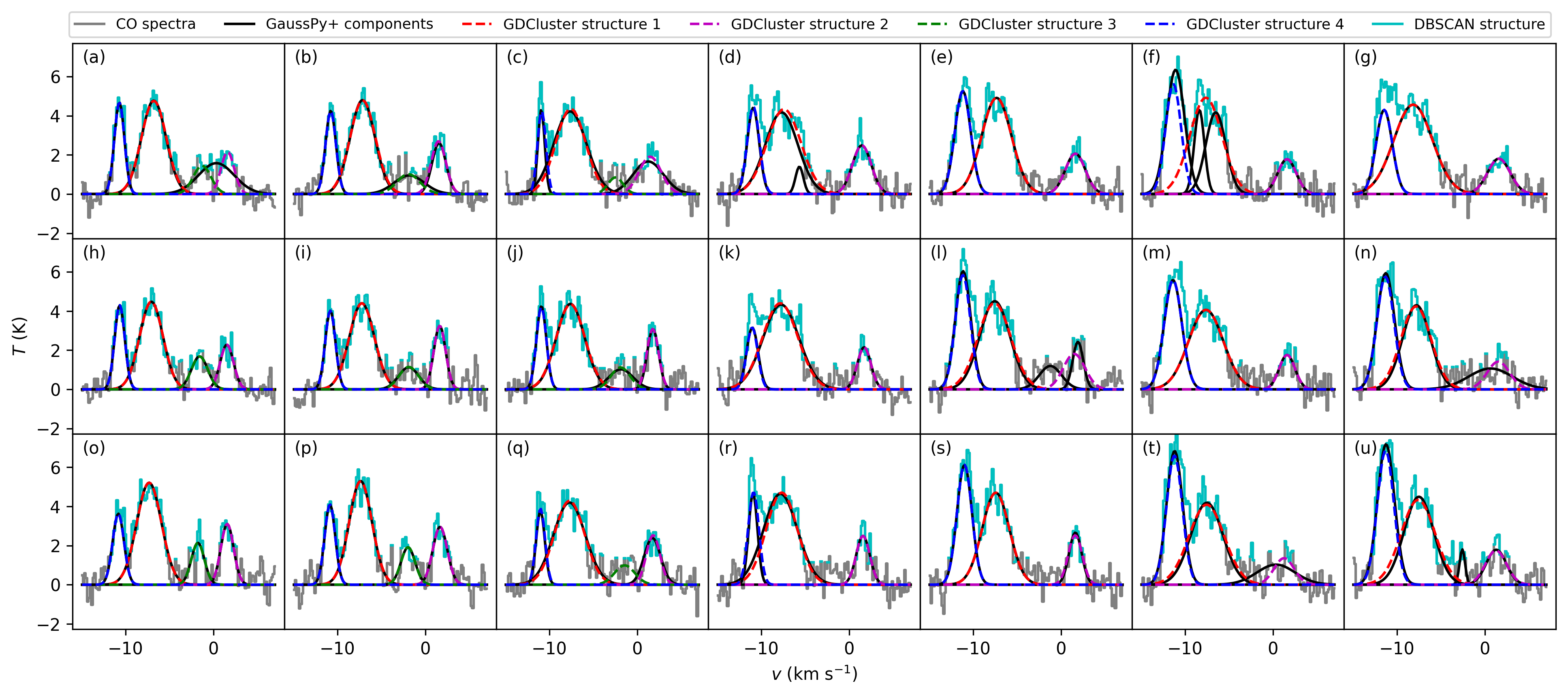}
    \caption{Spectra and decomposition results for several adjacent LoSs. Gray solid lines: observed spectra; cyan solid line: single DBSCAN structure; black solid line: Gaussian components decomposed by GaussPy+; dashed lines: four different GDCluster structures distinguished by different colors.}
    \label{fig::one_dbscan_multi_GDCluster}
\end{figure*}

If the spectral lines of a DBSCAN structure encompass \emph{any} fluxes of a GDCluster structure, we define this DBSCAN structure as \emph{covering} the GDCluster structure. Figure \ref{fig::one_cover_multi} presents histograms of (a) GDCluster structures covered by individual DBSCAN structures and (b) DBSCAN structures covered by individual GDCluster structures.
The largest DBSCAN structure mentioned earlier, spanning $\sim\SI{111}{deg^2}$, covers \num[group-separator={,}]{6549} GDCluster structures.
Figure \ref{fig::one_dbscan_multi_GDCluster} displays spectra and corresponding decomposition results (here we also present the decomposition results of GaussPy+) containing the largest DBSCAN structure.
Visual inspection suggests that this region contains at least four velocity components. 
When they are too close relative to their line-widths, their spectral lines blend and become inseparable in position-position-velocity (PPV) space, causing the multiple velocity components to be identified as a single structure by the DBSCAN algorithm.
Figure \ref{fig::one_cover_multi}(a) shows six GDCluster structures each covering more than nine DBSCAN structures.
These GDCluster structures all cover the largest DBSCAN structure along with faint structures separated from it due to their low intensities.
As shown in Figure \ref{fig::one_dbscan_multi_GDCluster}, GaussPy+ treated the emission near $v=\SI{-8}{\kms}$ as a single Gaussian component in all LoSs except for panel (f), where it split the emission into two components.
In contrast, GDCluster consistently interpreted the emission in this velocity range as originating from a single structure, which exhibits better spatial continuity.

\begin{figure}[ht!]
    \centering
    \plotone{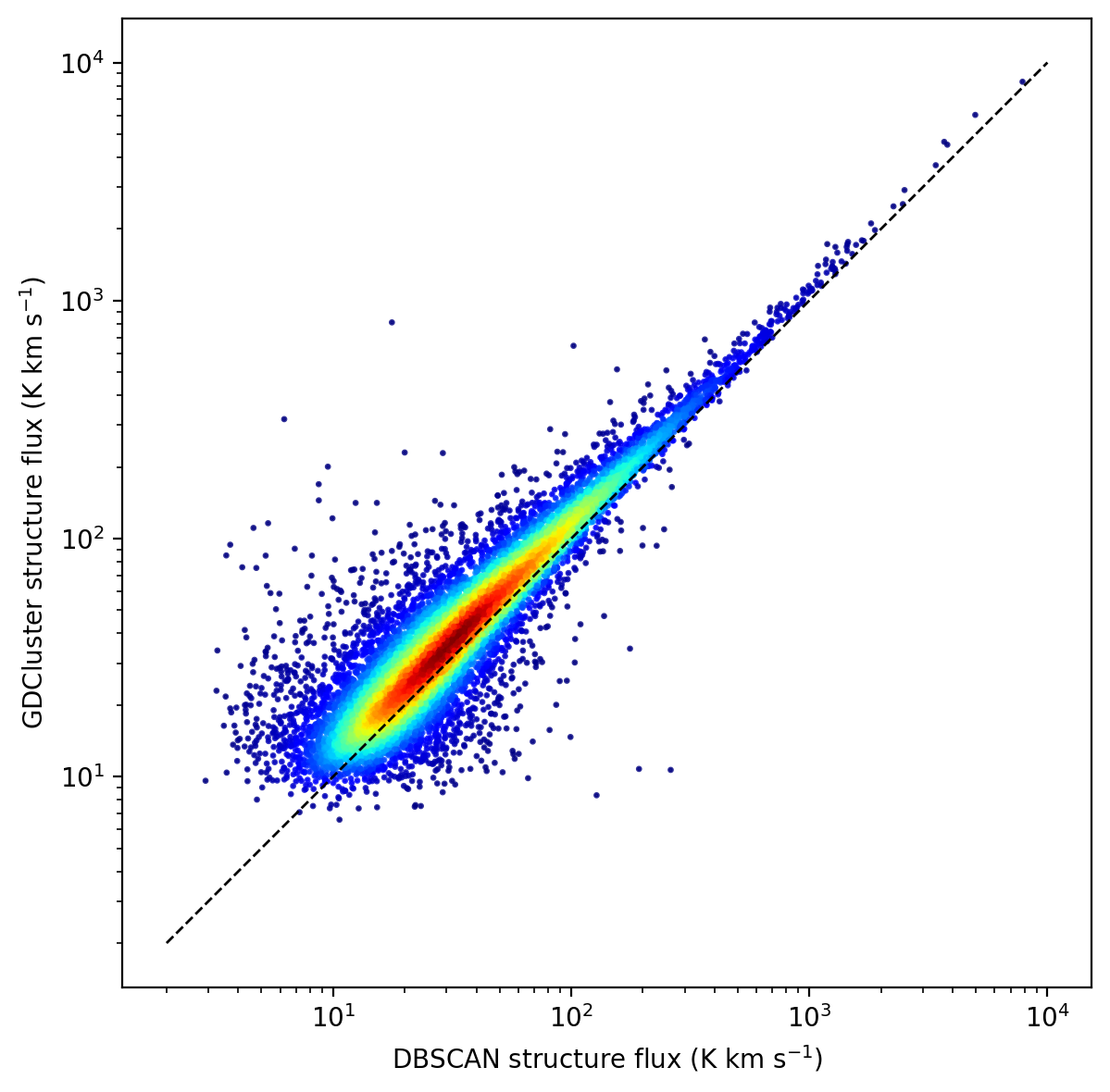}
    \caption{Density map of the flux of DBSCAN and GDCluster structures. Red indicates denser regions; the black dashed line marks the 1:1 flux relationship.}
    \label{fig::one_on_one}
\end{figure}

Among cross-matches between the two catalogs, \num[group-separator={,}]{10773} structures exhibit one-to-one correspondence. Figure \ref{fig::one_on_one} shows their flux distribution.
These one-to-one structures generally show linearly correlated fluxes, though most GDCluster structures have higher fluxes than their DBSCAN counterparts. Since Gaussian fitting retains all spectral line wings without noise cutoff—unlike DBSCAN's clipping—this systematic difference emerges.

\subsubsection{Performance comparison with Mock Data}
\begin{figure}[ht!]
    \centering
    \plotone{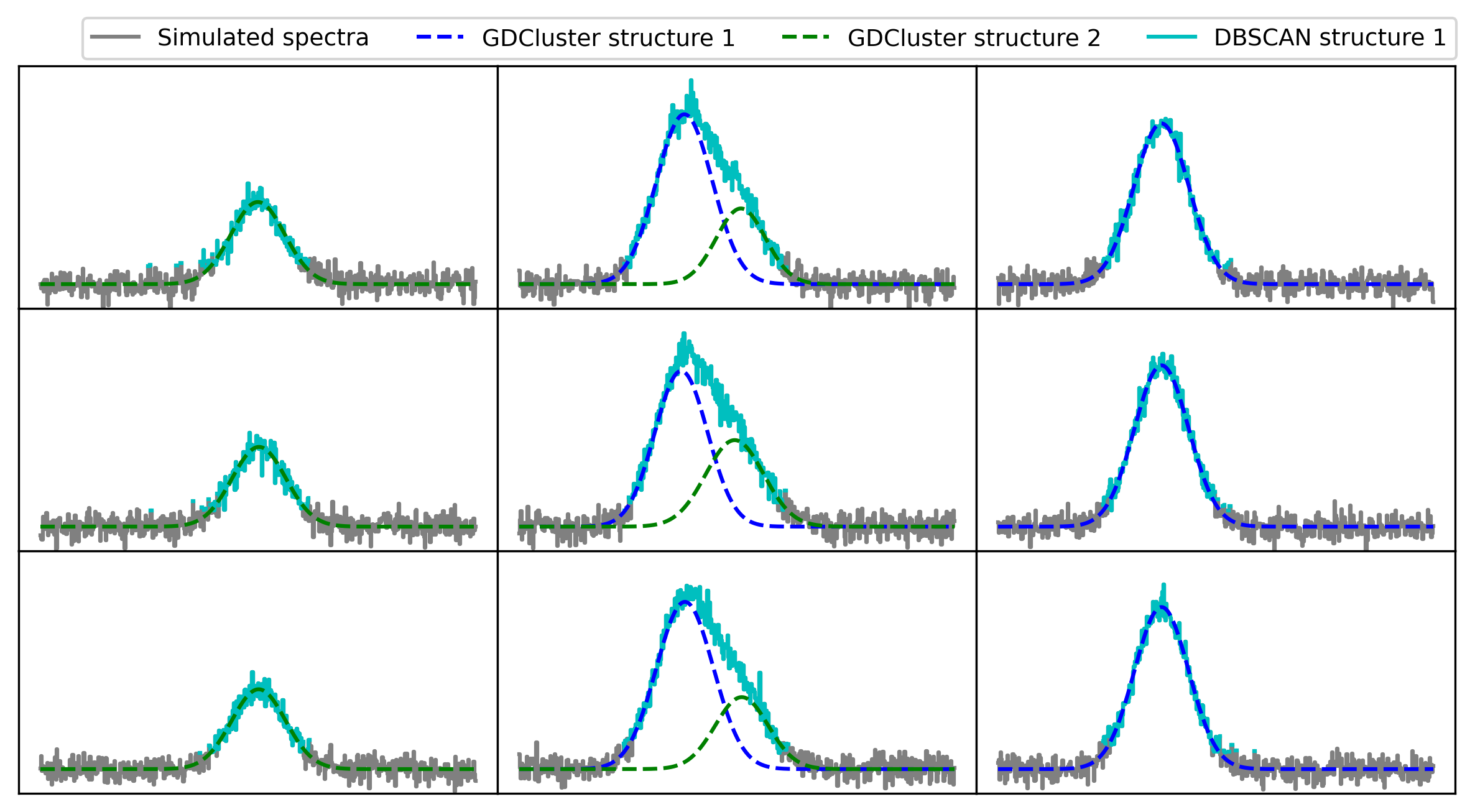}
    \caption{Simulated spectra and decomposition results for nine sightlines. Gray: simulated; blue/green dashed: GDCluster components; cyan: single DBSCAN structure.}
    \label{fig::spec_compare_2}
\end{figure}

\begin{figure}[ht!]
    \centering
    \plotone{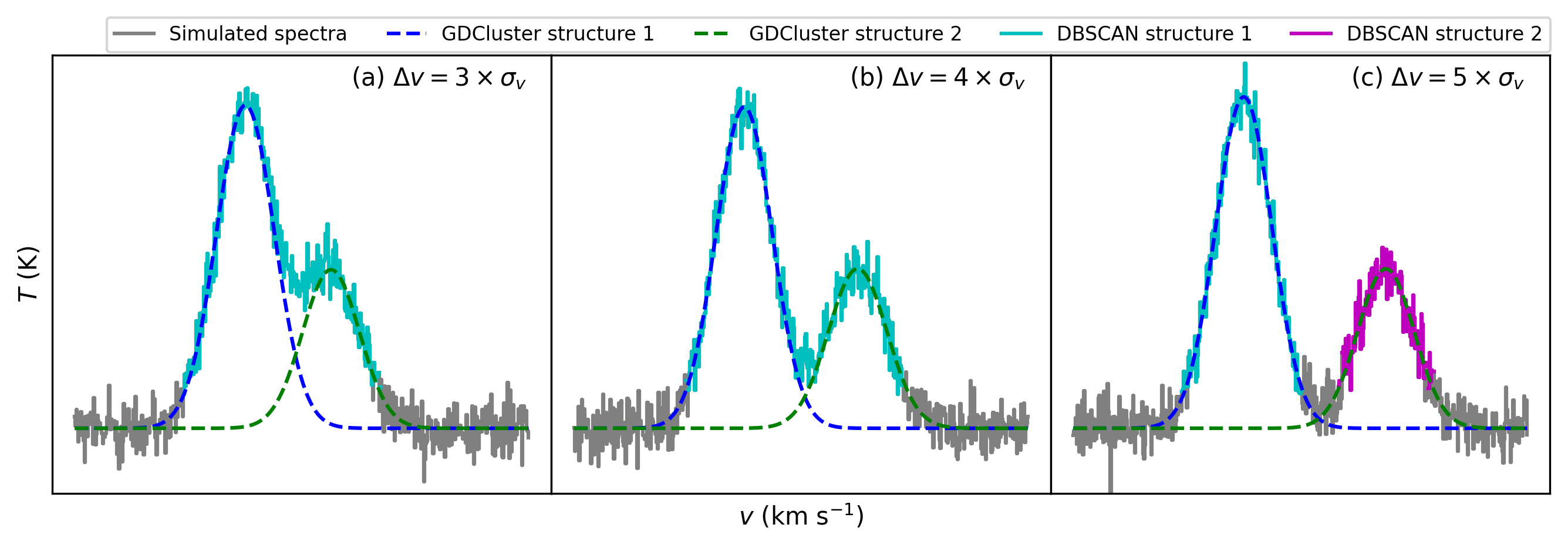}
    \caption{Simulated spectra and decomposition results for $\Delta v =$ (a) 3$\sigma_v$, (b) 4$\sigma_v$, (c) 5$\sigma_v$. Line styles match Figure \ref{fig::spec_compare_2}; magenta: additional DBSCAN structure.}
    \label{fig::vari_sigma}
\end{figure}

We evaluated three algorithms on mock data containing PPV-blended structures (common at low Galactic latitudes).
The spectral decomposition results of GDCluster and GaussPy+ are nearly identical; therefore, for simplicity, we present only the results from GDCluster.
Decomposition results appear in Figure \ref{fig::spec_compare_2}.
The data comprise two spatially intersecting structures with identical Gaussian components at every LoS.
The components have peak intensities of 20$\times$ and 10$\times$ the noise level, identical line-width of $\sigma_v=\SI{5}{\kms}$, and a velocity separation of $\Delta v=\SI{10}{\kms}$.
This configuration results in line blending along the central lines of sight (Figure \ref{fig::spec_compare_2}, column 2), which GDCluster resolves but DBSCAN merges.
Reduced blending (increased $\Delta v$) reveals spectral dips (Figure \ref{fig::vari_sigma}).
When dip intensity drops below the noise threshold, as shown in Figure \ref{fig::vari_sigma}(c), DBSCAN separates the structures.

In summary, GDCluster demonstrates superior capability for decomposing blended velocity structures, and the spectral decomposition results exhibit better spatial continuity.

\section{Discussion}
\label{sec::discussion}
\subsection{Limitations of GDCluster}
\label{sec::limitations}
One assumption of GDCluster is that the spectra of emitting gas exhibit multi-Gaussian profiles.
The spectral lines of interstellar gas, particularly those at low Galactic latitudes, often exhibit multiple velocity components along the LoS.
Line broadening mechanisms can obscure velocity gaps, obscuring the separation of these components.
A common approach to reconstructing their distinctions is to model line-centers and -widths.
The multi-Gaussian model for \HI 21\,cm emission lines has been studied for several decades and is generally valid for optically thin \HI in most sources \citep{2018A&A...619A..58K}.
\HI absorption line surveys have shown that Gaussian models can accurately reproduce both opacity and emission profiles \citep{2003ApJS..145..329H}, as well as the spin temperature of the CNM predicted by analytical methods \citep{2015ApJ...804...89M}. 
\citet{2003ApJ...593..831M} pointed out that, as the kinetic temperature increases, the spectrum becomes closer to Gaussianity.

However, for optically thick lines such as $\ce{^{12}CO}\quad J=1$--$0$, the multi-Gaussian assumption must be applied with caution.
In the presence of noise, a flat-topped opaque line becomes nearly indistinguishable from overlapping emissions of several kinematically close Gaussians. In other words, optically thick lines can be approximated by several Gaussians, but further revision is necessary when interpreting gas properties.

Another phenomenon that challenges the multi-Gaussian assumption is HISA.
What HISA actually violates is the ``emission line profile" assumption, which breaks down when broad emission from the WNM is absorbed by foreground CNM.
The lack of distance information prevents us from fitting absorbing CNM components prior to HISA identification.
Instead, we approximate \HI spectra containing potential HISA features using multiple Gaussian emission components, identify HISA candidates, and subsequently model them with radiative transfer.

\subsection{Hyper-parameters and Efficiency of GDCluster}
It is important to note that, due to noise and limited resolution, a given spectrum may have multiple viable best-fit solutions. Although spatial information reduces this degeneracy, decomposition results can still vary slightly depending on the hyper-parameter settings. Among these, the smoothing factor $\sigma_{\mathrm{min}}$, the SNR threshold, and the ratio of line-center difference to line-width have the most significant influence.

For the $\ce{^{12}CO}$ dataset with a $5461\times1261\times883$ PPV cube, the full decomposition required approximately 160 CPU hours (AMD EPYC 7302 with a clock rate of \SI{3}{GHz})—considerably faster than many existing algorithms, such as DBSCAN or the combination of GaussPy+ decomposition with Acorns clustering adopted by \citet{2024AJ....167..220Z}.
The efficiency of our proposed algorithm enables decomposition using a range of hyper-parameters, allowing us to select the optimal result.

\subsection{Definition of Clouds}
\label{sec::def_of_clouds}
The definition of molecular clouds fundamentally influences derived properties and statistical characterization.
Visual identification typically defines clouds as ``topologically closed surfaces of antenna temperature in PPV space" \citep{1987ApJ...319..730S,2001ApJ...562..348O}. Following this traditional paradigm, numerous identification methods operate directly in PPV space, such as Clumpfind \citep{1994ApJ...428..693W}, Dendrograms \citep{2008ApJ...679.1338R}, Fellwalker \citep{2015A&C....10...22B}, SCIMES \citep{2015MNRAS.454.2067C}, DBSCAN \citep{2020ApJ...898...80Y}, and etc.

Conversely, decomposition-based algorithms (typically Gaussian) process the spectral dimension first.
These define clouds as density enhancements \citep{2008A&A...483..461H} or clusters \citep{2013A&A...554A..55H,2019MNRAS.485.2457H,2020ApJ...902..120P} in parameter space.
This definitional divergence stems from differential spectral treatment: non-decomposition methods handle spatial and spectral dimensions nearly equivalently, while spectral decomposition applies specialized processing to velocity information.

When peculiar motions are negligible and velocity serves as a proxy for distance, emissions in PPV space can be mapped to position-position-position (PPP) space via Galactic rotation models. In this idealized case, continuous spectra correspond to continuous radial gas distributions.
However, blended spectral lines from multiple structures do not necessarily indicate spatial overlap. 
Furthermore, velocity perturbations injected by stellar feedback, gravitational collapse, and supernovae exacerbate velocity blending. Crucially, \citet{2022ApJ...925..201P} reveal that gas can extend more than \SI{3}{kpc} in the radial distance but exhibit similar radial velocity; \citet{2024AJ....167..220Z} demonstrate that clouds in the Cygnus X North region with similar radial velocities are physically separated by $\sim\SI{700}{pc}$.
These complexities necessitate specialized spectral treatment when multiple clouds present in the same LoS.

When modeling extended gas with similar physical properties using Gaussian components, emissions can be characterized by line-centers and -widths. Distinct structures become resolvable when their line-center separations exceed typical line-widths, as shown in Figure \ref{fig::vari_sigma}.
Consequently, algorithms employing spectral decomposition prove significantly more effective at resolving cloud structures in kinematically complex LoSs than non-decomposition approaches.

\section{Summary}
\label{sec::summary}
In this work, we propose a fully automatic algorithm, GDCluster, to decompose interstellar gas into structures.
The decomposition process consists of three main stages:  
1) accurate estimation of multi-Gaussian initial parameters via derivative spectroscopy;
2) constrained multi-Gaussian fitting informed by spatial-continuity;
3) clustering based on spatial-continuity constraints and uncertainties.

Benefiting from improvements on derivative spectroscopy technique, GDCluster resolves overlapping Gaussian profiles with near-optimal initial parameters, accelerating the fitting process.
Spatial-continuity constraints suppress noise-induced peaks, adjust Gaussian counts, and constrain parameters, improving clustering validity.
A visually inspired clustering criterion minimizes hyper-parameter tuning while preserving interpretability, enabling decomposition of large-scale, even all-sky, spectral line surveys into gas structures.

Finally, we compared the performance of GDCluster, DBSCAN, and GaussPy+ algorithm both on the MWISP survey and mock data.
Results show that GDCluster is more effective than DBSCAN in separating regions containing multiple velocity components, and the decomposition results exhibit better spatial continuity than GaussPy+.
Specifically, the largest DBSCAN structure, covering $\sim\SI{111}{deg^2}$ and spanning $\sim\SI{43.7}{\kms}$ with obvious multiple velocity component, are decomposed into more than 6500 structures by GDCluster.
The decomposition results of mock data clearly shows that DBSCAN can resolve two Gaussians only if the spectral dip between them drops below the noise cutoff, whereas GDCluster can reconstruct their line-centers and -widths, facilitating their separation.
Since GDCluster does not truncate spectra, the resulting structures recover more flux than those obtained with DBSCAN.

Considering the limitations of GDCluster discussed in Section~\ref{sec::limitations}, we identify two directions for improvement.
First, we plan to introduce the opacity as a fitting parameter to better reconstruct optically thick spectra.
Second, we aim to replace the multi-Gaussian model with a radiative transfer model to fit HISA features.
If molecular gas is well mixed with cold atomic gas, rotational transitions of molecules can serve as efficient coolants, producing \HI Narrow Self-Absorption (HINSA) \citep{2003ApJ...585..823L}.
In this context, \ce{CO} structures can serve as indicators of foreground CNM structures, enabling the radiative transfer model to fit HINSA features.
We have already made progress in this direction, and detailed results will be presented in forthcoming publications.

\begin{acknowledgments}
We would like to thank the anonymous referee for providing constructive comments that helped improve the quality of this paper.
We are grateful to Yang Su and Shiyu Zhang for their careful review and insightful feedback on the results of the proposed algorithm.
We thank Keping Qiu, Zhiyu Zhang, Hongchi Wang, Zhibo Jiang, Xuepeng Chen, Min Fang, Yan Sun, Qingzeng Yan, Lixia Yuan, Ji Yang, Miaomiao Zhang, Yuehui Ma, Xin Zhou for their valuable suggestions.
This work is supported by the National Natural Science Foundation of China with grant 12041305 and the National Key R\&D Program of China with grant 2023YFA1608000.
This research made use of the data from MWISP project, which is a multi-line survey in $\ce{^{12}CO}$/$\ce{^{13}CO}$/$\ce{C^{18}O}$ along the northern Galactic plane with the PMO 13.7 m telescope.
The work makes use of publicly released data from the HI4PI survey that combines the EBHIS in the Northern Hemisphere with the GASS in the Southern Hemisphere.
This work made use of the data from FAST. FAST is a Chinese national mega-science facility, operated by National Astronomical Observatories, Chinese Academy of Sciences.

\end{acknowledgments}

%

\vspace{5mm}


\software{astropy \citep{2013A&A...558A..33A,2018AJ....156..123A}}
\facility{PMO:DLH}



\appendix

\section{A parameter-free noise estimation method for spectral data}
\label{apx::noise_esti}
Suppose a spectrum  $I(c)$ is sampled across $N$ channels with additive white Gaussian noise (AWGN). In practice, the derivative is approximated using the forward difference:
    \begin{equation}
    	I'(c)\approx\frac{I(c+h)-I(c)}{h},
    	\label{equ::foreward_diff}
    \end{equation}
    in which case the $n$th-order derivative of a spectrum can be expressed as 
    $$I^{(n)}\approx\sum_{k=0}^{n}(-1)^{k}\binom{n}{k}I_{n-k}.$$
    Using the rule of error propagation, the noise rms at the $n$th-order derivative is given by:
    \begin{equation} 
    	\alpha_{n} = \alpha_{0}\cdot\sqrt{\sum_{k=0}^{n}\left[(-1)^{k}\binom{n}{k}\right]^{2}} = \alpha_{0} \cdot \sqrt{\sum_{k=0}^{n} \left( \frac{n!}{(n-k)!k!} \right)^{2}},
    	\label{equ::error_prop}
    \end{equation}
    where $\alpha_{0}$ and $\alpha_{n}$ represent the standard deviations of the 0th- (original spectrum) and $n$th-order derivatives, respectively. Since Equation \ref{equ::foreward_diff} reduces the number of independent channels by half, Equation \ref{equ::error_prop} holds when $n \le m = \lfloor \log_{2}(N) \rfloor$. To suppress the signal as much as possible and reduce the impact of outliers from narrow signals, the Median Absolute Deviation (MAD) is used to estimate the variability of the $m$th-order derivative:
    \begin{equation}
    	\mathrm{MAD}=\mathrm{median}(|X_{i}-\tilde{X}|),\,\tilde{X}=\mathrm{median(X)}.            
    	\label{equ::mad}
    \end{equation}
    For normally distributed AWGN, the MAD is related to the standard deviation by $\alpha\approx 1.4826\cdot\mathrm{MAD}.$ Finally, the noise rms in the original spectrum is calculated as:
    \begin{equation}
    	\alpha_{0} = \frac{\mathrm{MAD}[I^{(\alpha)}(v)]}{0.6745 \cdot \sqrt{\sum_{k=0}^{n} \left( \frac{n!}{(n-k)!k!} \right)^{2}}}.
    \end{equation}

\section{Decomposition results of $^{13}$CO and C$^{18}$O}
\label{apx::stats_13CO_C18O}

\begin{figure*}[ht!]
    \plotone{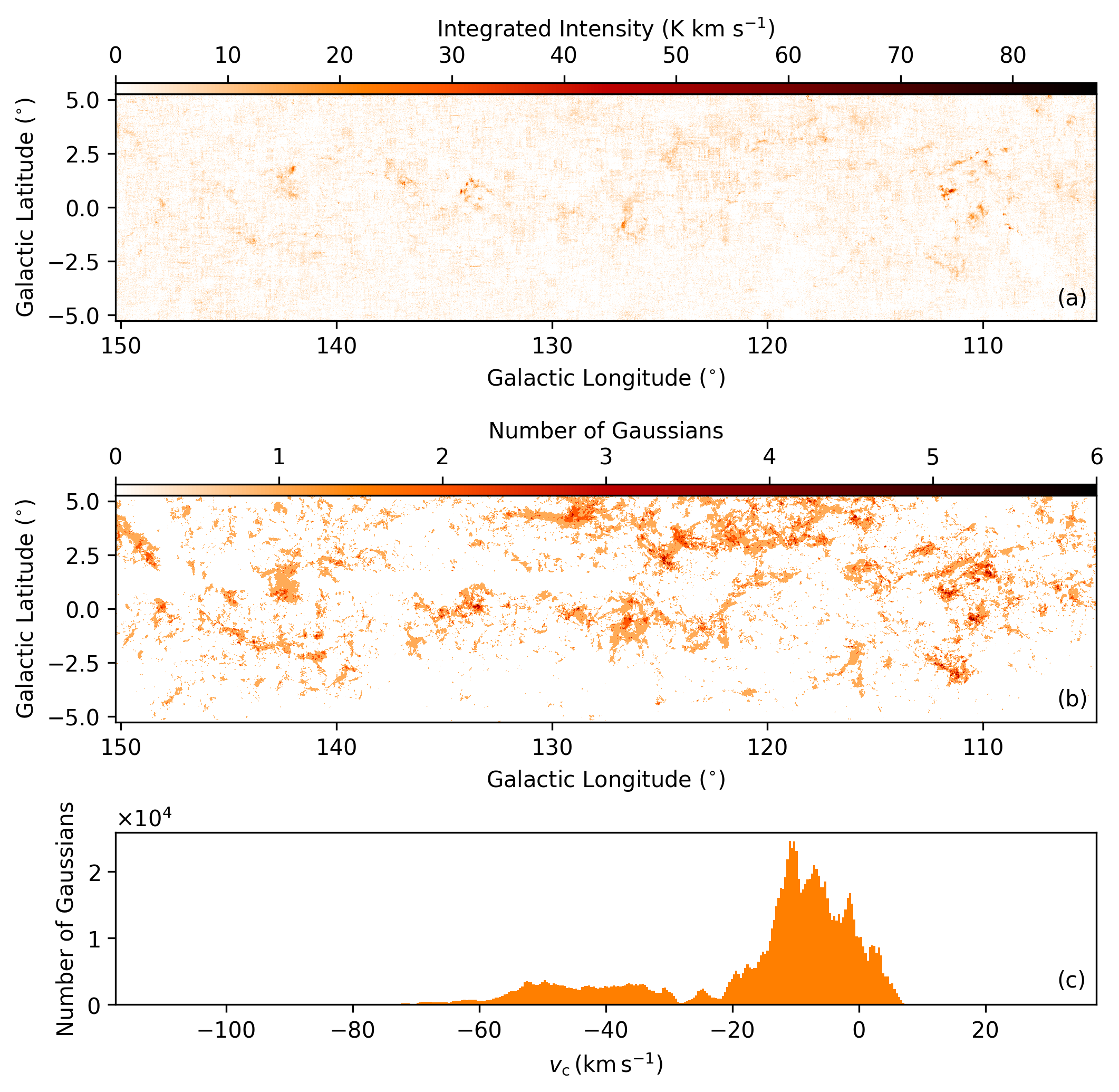}
    \caption{The same as Figure \ref{fig::m0_and_comp_num_in_lb_HI4PI_mol} for $^{13}\text{CO}$ data.}
\end{figure*}

\begin{figure*}[ht!]
    \plotone{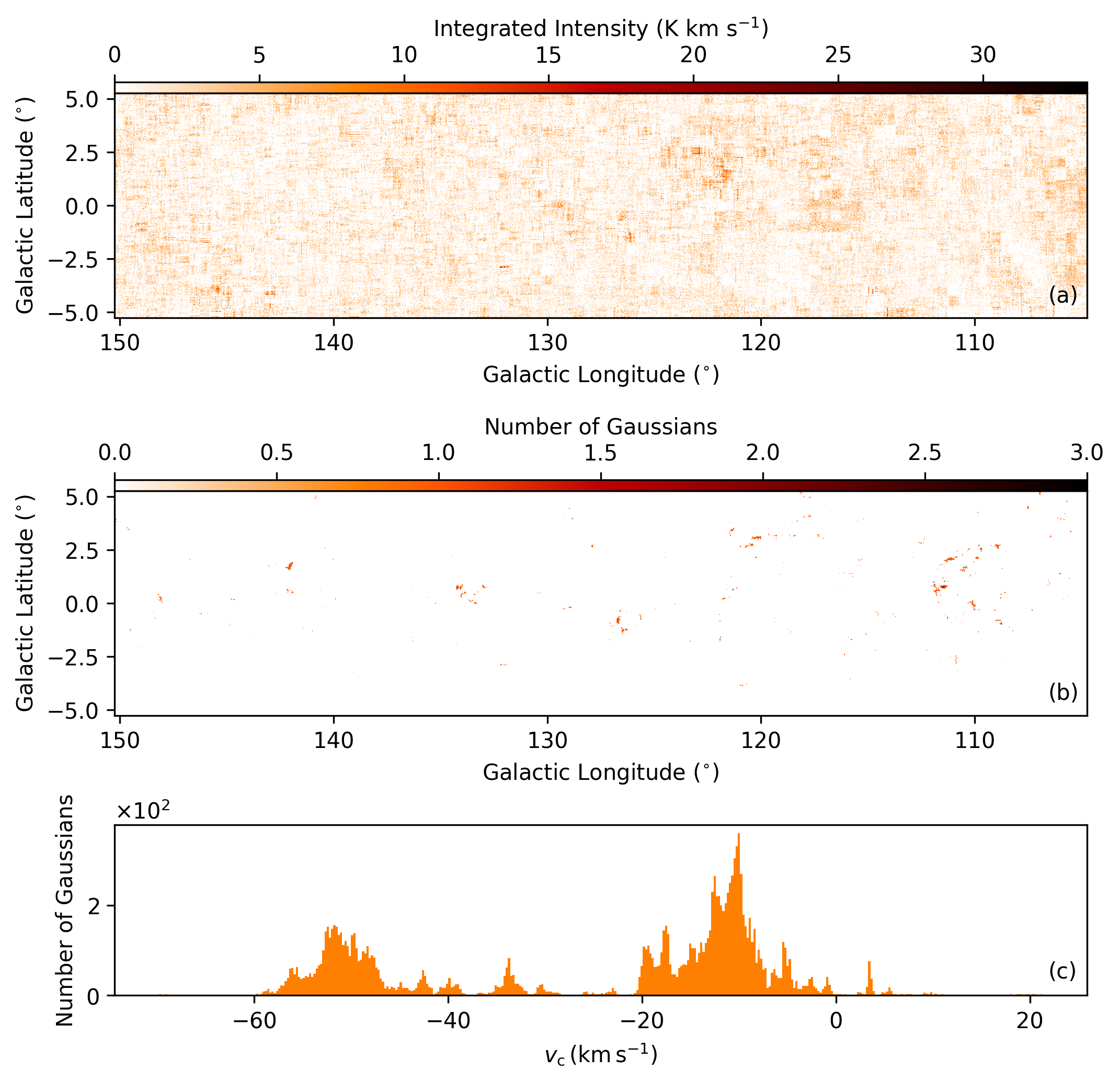}
    \caption{The same as Figure \ref{fig::m0_and_comp_num_in_lb_HI4PI_mol} for $\text{C}^{18}\text{O}$ data. The intensity is integrated over the entire velocity range, resulting in a noise-like moment 0 map.}
\end{figure*}

\begin{figure*}[ht!]
	\gridline{\fig{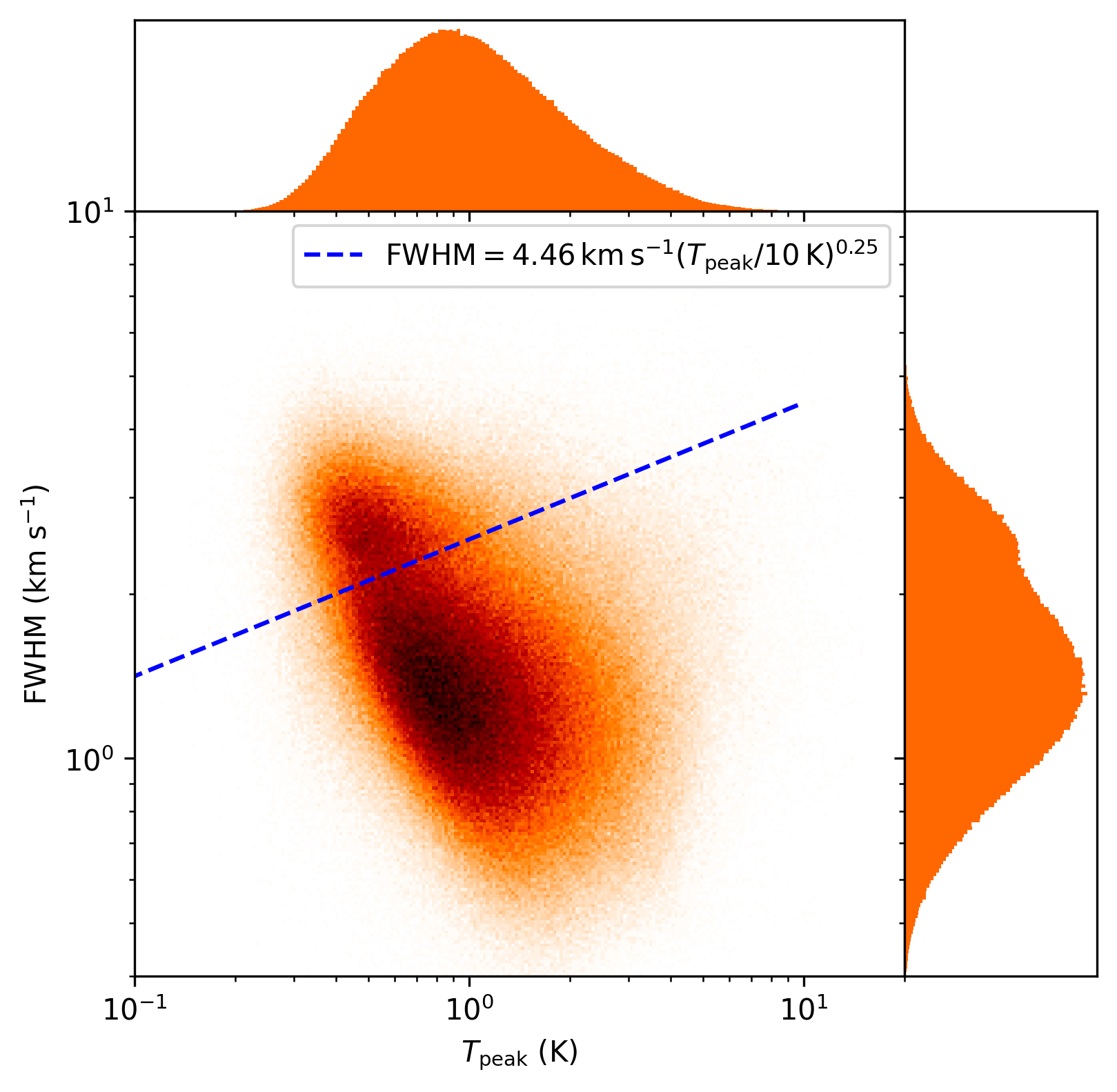}{0.45\textwidth}{(a)}
			  \fig{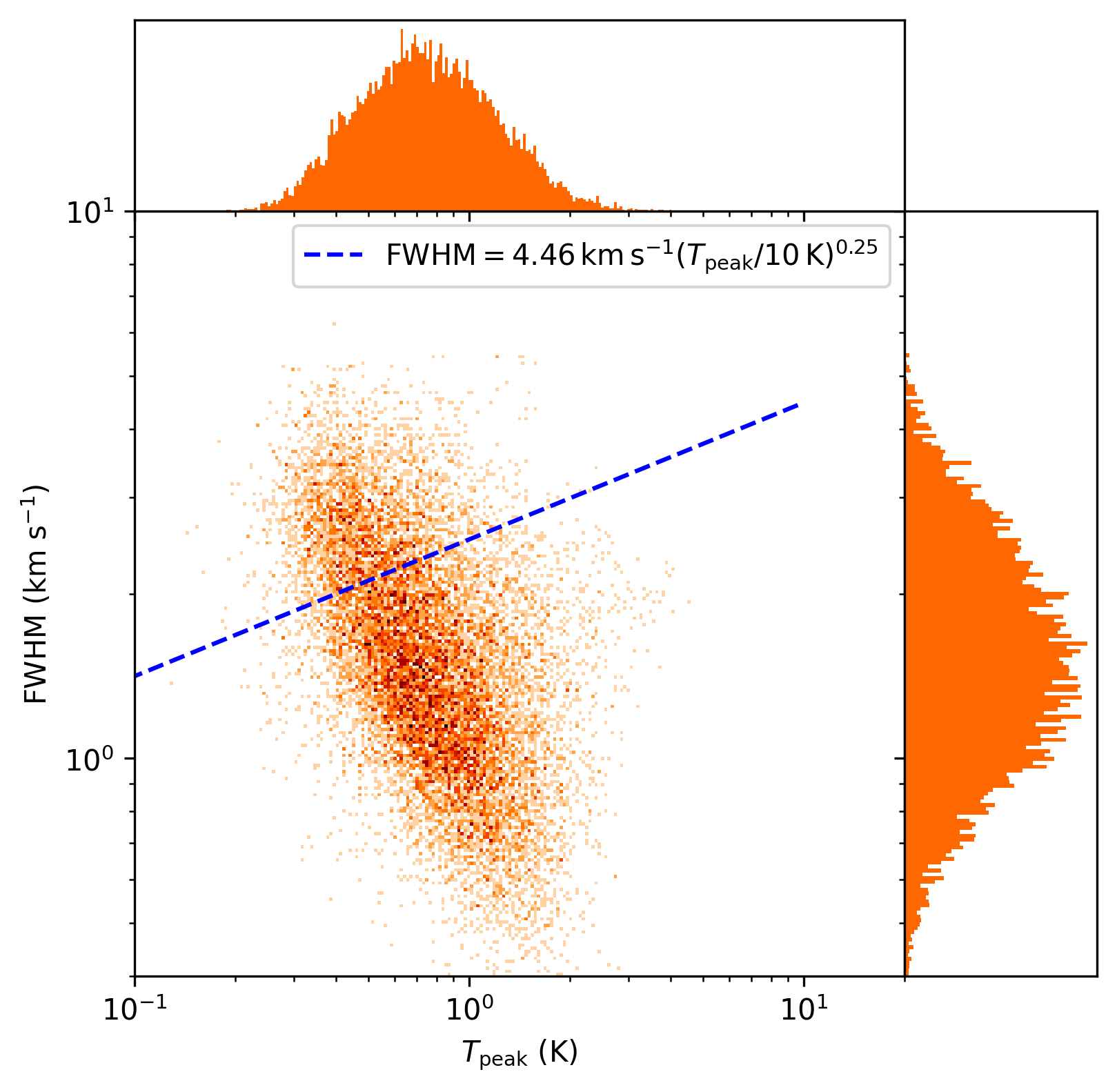}{0.45\textwidth}{(b)}}
	\caption{The same as Figure \ref{fig::comp_wise_T_w_relation_HI4PI_mol} for all (a) $^{13}\text{CO}$ and (b) $\text{C}^{18}\text{O}$ Gaussians.}
\end{figure*}


\bibliography{sample631}{}
\bibliographystyle{aasjournal}



\end{document}